\documentclass[pdflatex,sn-mathphys-num]{sn-jnl}

\usepackage{graphicx}%
\usepackage{multirow}%
\usepackage{amsmath,amssymb,amsfonts}%
\usepackage{amsthm}%
\usepackage{mathrsfs}%
\usepackage[title]{appendix}%
\usepackage{xcolor}%
\usepackage{textcomp}%
\usepackage{manyfoot}%
\usepackage{booktabs}%
\usepackage{algorithm}%
\usepackage{algorithmicx}%
\usepackage{algpseudocode}%
\usepackage{listings}%
\usepackage{subcaption}
\usepackage[T1]{fontenc}
\usepackage{lmodern}
\usepackage{caption}
\usepackage{mathrsfs}
\usepackage{natbib}

\begin{document}

\title[Article Title]{Bayesian uncertainty analysis for underwater 3D reconstruction with neural radiance fields}

\author[1,2]{\fnm{Haojie} \sur{Lian}}
\author[1]{\fnm{Xinhao} \sur{Li}}
\author*[3,4,5]{\fnm{Yilin} \sur{Qu}}\email{quyilin\_95@126.com (Y.Q.)}
\author[6]{\fnm{Jing} \sur{Du}}
\author[6]{\fnm{Zhuxuan} \sur{Meng}}
\author[7]{\fnm{Jie} \sur{Liu}}
\author*[2]{\fnm{Leilei} \sur{Chen}}\email{chenllei@mail.ustc.edu.cn (L.C.)}

\affil[1]{\orgdiv{Key Laboratory of In-situ Property-improving Mining of Ministry of Education}, \orgname{Taiyuan University of Technology}, \orgaddress{\city{Taiyuan}, \postcode{100190}, \country{China}}}
\affil[2]{\orgdiv{Henan International Joint Laboratory of Structural Mechanics and Computational Simulation, College of Architecture and Civil Engineering}, \orgname{Huanghuai University}, \orgaddress{\city{Zhumadian}, \postcode{463000}, \country{China}}}
\affil[3]{\orgdiv{School of Marine Science and Technology}, \orgname{Northwestern Polytechnical University}, \orgaddress{\city{Xi’an}, \postcode{710072}, \country{China}}}
\affil[4]{\orgdiv{Unmanned Vehicle Innovation Center}, \orgname{Ningbo Institute of Northwestern Polytechnical University}, \orgaddress{\city{Ningbo}, \postcode{315103}, \country{China}}}
\affil[5]{\orgdiv{Key Laboratory of Unmanned Underwater Vehicle Technology of Ministry of Industry and Information Technology}, \orgname{Northwestern Polytechnical University}, \orgaddress{\city{Xi’an}, \postcode{710072}, \country{China}}}
\affil[6]{\orgdiv{Center for Strategic Assessment and Consulting}, \orgname{Academy of Military Science}, \orgaddress{\city{Beijing}, \postcode{100091}, \country{China}}}
\affil[7]{\orgdiv{Computer Engineering Department}, \orgname{Taiyuan Institute of Technology}, \orgaddress{\city{Taiyuan}, \postcode{030008}, \country{China}}}

\abstract{Neural radiance fields (NeRFs) are a deep learning technique that can generate novel views of 3D scenes using sparse 2D images from different viewing directions and camera poses. As an extension of conventional NeRFs in underwater environment, where light can get absorbed and scattered by water, SeaThru-NeRF was proposed to separate the clean appearance and geometric structure of underwater scene from the effects of the scattering medium. Since the quality of the appearance and structure of underwater scenes is crucial for downstream tasks such as underwater infrastructure inspection, the reliability of the 3D reconstruction model should be considered and evaluated. Nonetheless, owing to the lack of ability to quantify uncertainty in 3D reconstruction of underwater scenes under natural ambient illumination, the practical deployment of NeRFs in unmanned autonomous underwater navigation is limited. To address this issue, we introduce a spatial perturbation field $\mathcal{D}_{\boldsymbol{\omega}}$ based on Bayes’ rays in SeaThru-NeRF and perform Laplace approximation to obtain a Gaussian distribution $\mathcal{N}(0,\Sigma)$ of the parameters $\boldsymbol{\omega}$, where the diagonal elements of $\Sigma$ correspond to the uncertainty at each spatial location. We also employ a simple thresholding method to remove artifacts from the rendered results of underwater scenes. Numerical experiments are provided to demonstrate the effectiveness of this approach.}


\maketitle

\section{Introduction}

3D reconstruction of underwater scenes is an important and challenging research topic in marine environmental research \citep{Zhang1,Zhang2,Zhang3}. Traditional 3D reconstruction methods use geometric constraints and certain prior assumptions to recover the structure of target objects in a discrete 3D space, such as multi-view stereo (MVS), structure from motion (SfM) and voxel grid. Due to reliance upon discrete representations, the aforementioned methods have difficulty recovering detailed geometric information of complex structures. In addition, these methods have a high requirement of the the quality and quantity of the input data, resulting in a cumbersome data acquisition and pre-processing procedure.

In recent years, Neural Radiance Fields (NeRFs) \citep{nerf2021}  have emerged as novel approach for reconstructing 3D representation of a scene from 2D images. NeRFs fall into the category of neural rendering, which combine classical ray tracing methods in computer graphics and deep learning techniques. The key idea of NeRFs is to make a fully connected neural network learn a continuous function that maps 3D spatial coordinates and viewing directions to colors and optical densities of scenes. After the neural networks are trained using a sparse set of input 2D images that are collected from different viewing angles and camera pose, NeRFs are able to synthesize novel views of a scene from arbitrary viewpoints. Compared to traditional methods, NeRFs not only enhance image quality of 3D reconstruction, but more importantly, improve the efficiency considerably because it is simple to use and only require a sparse number of input views.

However, the presence of complex optical effects such as backscatter and attenuation in underwater environments poses great challenges for application of NeRFs. In shallow water with natural ambient illumination, the propagation of light follows a very different physical law from that in air, leading to a significant degradation in the reconstruction quality of NeRFs. To address this challenge, SeaThru-NeRF \citep{seathrunerf} extends the NeRFs for the first time to scattering media such as underwater. It is able to synthesise realistic renderings of novel views, while isolating clean scene appearance and geometry from the effects of scattering media. This not only helps to extend the application scenarios of NeRFs, but also helps to recover the hidden scene details from the scattering medium data, which is of great value in the fields of autonomous underwater vehicles (AUVs) navigation, and unmanned vehicles under severe weather \citep{lian2023lidar}.

Despite these advancements, the complexity of underwater environment introduces inherent uncertainty in modeling optical effects. However, most existing works treat NeRFs as a deterministic model, ignoring its inherent sources of uncertainty. In fact, under different environments and data distributions, NeRFs exhibits certain uncertainty and variability in reconstruction accuracy and geometric detail restoration, which seriously impacts the model's generalisation and robustness. This directly affects the visual quality of AUVs during ocean exploration and navigation, thereby introducing unforeseen failure risks to probability-bound downstream tasks \citep{estimate-uncertainty-field} such as underwater inspection and monitoring \citep{terracciano2020marine,ioannou2024underwater}, underwater navigation and localisation \citep{maurelli2022auv,martz2020survey}, and underwater infrastructure inspection \citep{halder2023robots}. Fortunately, through uncertainty quantification methods, we can enhance the reliability assessment of the model, thus improving the quality of AUVs decision-making in various real-world tasks and providing important guarantees for the successful completion of these tasks.

Uncertainty quantification plays a crucial role in reducing uncertainty in optimization and decision-making processes. In recent years, with the continuous advancement of deep learning techniques, various uncertainty quantification methods such as deep ensemble, variational inference, MC-dropout, and others have emerged as hot research topics in both academia and industry \citep{gawlikowski2023survey}. To enhance the robustness and reliability of NeRFs in practical applications, some studies have begun to combine these methods to explore the uncertainty of NeRFs. However, to the authors’ best knowledge, there is no existing literature that documents the uncertainty quantification related to underwater NeRFs, especially under natural ambient illumination.

\begin{figure}
	\centering
	\includegraphics[width=0.9\textwidth]{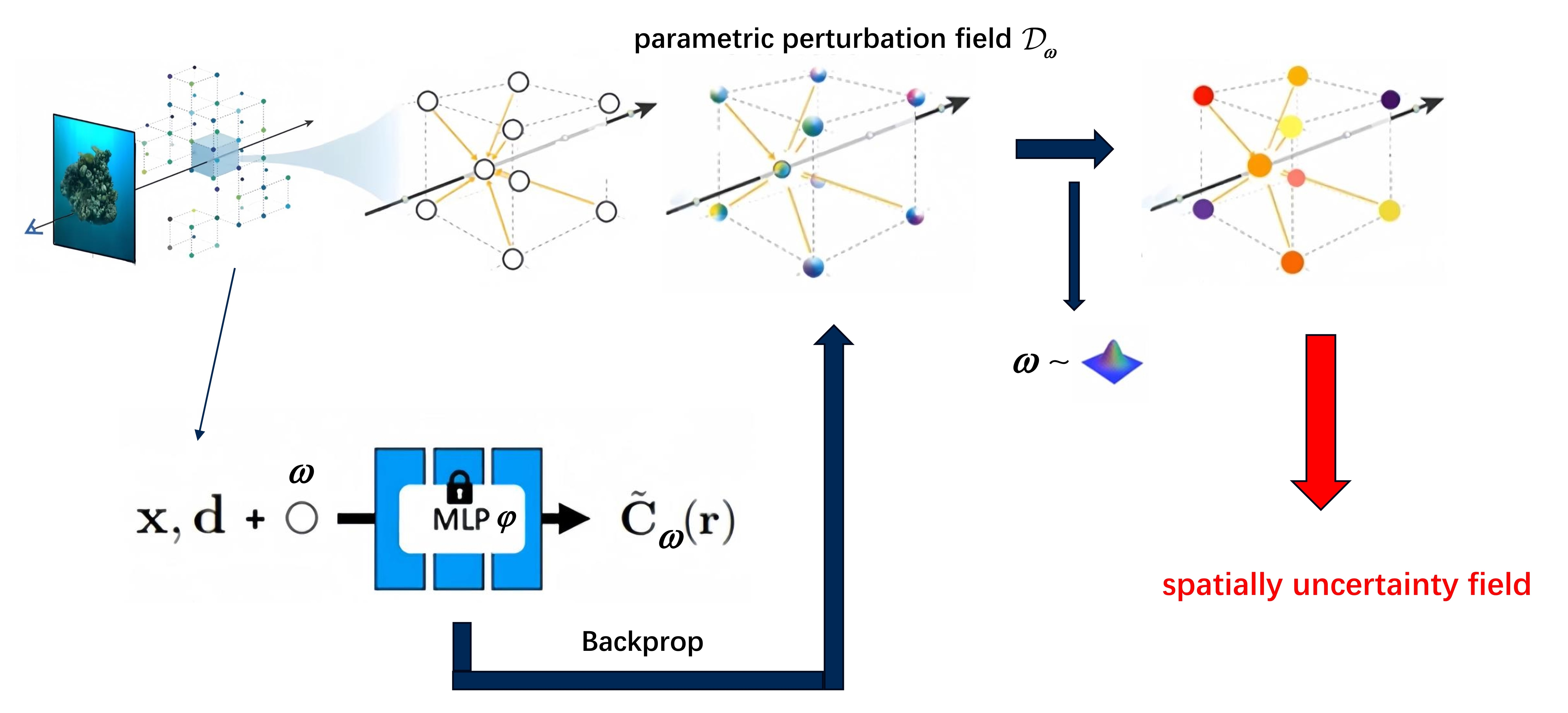}
	\caption{Pipeline.}
	\label{FIG:pipeline}
\end{figure}

To address the research gap mentioned above, we attempt to introduce a learnable spatial perturbation field $\mathcal{D}_{\boldsymbol{\omega}}$ and perform Laplace approximation based on Bayes’ rays \citep{bayes_rays}  to quantify the uncertainty of pre-trained SeaThru-NeRF model (Figure \ref{FIG:pipeline}). The perturbation field is used to perturb the input coordinate $\mathbf{x}$ of the original SeaThru-NeRF network, thus indirectly leading to the reparameterization of the entire model. By using the Laplace approximation method, the uncertainty of each spatial location is estimated based on the difference between the original reconstruction results and the perturbed results. In addition, we utilize a thresholding method to remove artifacts caused by occlusions or incomplete data.

In this work, we demonstrate that our model can explicitly infer spatial uncertainty in both synthetic and real-world underwater scenes. In summary, we makes the following key contributions:
\begin{itemize}
\item This work introduces uncertainty quantification into NeRFs for underwater scenes for the first time, allowing for the analysis of model reliability and enhancement of its robustness.
\item By introducing an additional perturbation field and performing a Laplace approximation, we avoid additional training, costly sampling, or accessing the training images.
\item Furthermore, the model can be used for post-processing to remove artifacts.
\end{itemize}

\section{Related works}
\subsection{NeRFs in underwater scenes}
NeRFs represents a novel approach leveraging deep learning techniques to reconstruct 3D scenes. However, because of the effects of backscatter and attenuation lead to the complex light propagation in underwater, makes accurate 3D scene reconstruction by traditional NeRFs difficult. To address this issue, researchers have proposed various improvements methods. WaterNeRF \citep{waternerf} uses NeRFs to additionally learn water column parameters, and combines them with light transport model \citep{light-transport} for color correction. By matching the corrected color distribution to the reference image distribution and using the Sinkhorn loss function \citep{sinkhorn} for training, it enables to produce the consistent color-corrected results from varying viewpoints during the inference stage. U2NeRF \citep{u2nerf} extends UPIFM \citep{UPIFM}. It modifies the generalizable NeRF transformer (GNT) \citep{GNT} to predict scene radiance, direct and back scatter transmission maps, and employs variational autoencoder (VAE) to predict the global background light component. By integrating these four components with the image formation model \citep{UPIFM}, it reconstructs the original underwater image. WaterHE-NeRF \citep{waterhenerf} designs a novel water-ray tracing field based on the Retinex model \citep{Retinex-model} to learn color, density, and illuminance attenuation. By controlling the intensity of illuminance attenuation, it can generate both degraded and restored multi-view images simultaneously. SeaThru-NeRF \citep{seathrunerf} based on the SeaThru \citep{Sea-thru}, replaces the traditional single opaque object density with the sum of the object density and the medium density. The final pixel color consists of the object radiance attenuated by the medium and the accumulated medium radiance along the light. It can not only generate realistic images from new viewpoints but also utilize the learned object and medium parameters to remove the effects of the scattering medium, reconstructing the true appearance and color of the scene as if the image were taken in the air.

\subsection{Uncertainty in deep learning}

In deep learning, uncertainty refers to the confidence level in the model's prediction results. Effective estimation and management of uncertainty can not only improve the model's predictive performance but also enhance its robustness to anomalies \citep{r.2.2.1,r.2.2.2}. Variational inference \citep{r.2.2.9,r.2.2.10,r.2.2.11} approximates the intractable posterior distribution by introducing a prespecified family of variational distribution and minimizes the evidence lower bound (ELBO) to reduce the discrepancy between them. While this approach provides comprehensive uncertainty estimates, it has high computational complexity. Deep ensemble strategies \citep{r.2.2.6,r.2.2.8} use multiple independently trained models combined with diverse data subsets, model structures, or initialisation methods to reduce the variance of a single model, thus improving generalisation and robustness. However, training multiple models requires more computational resources and storage space, and may introduce additional prediction delays. MC-dropout-based methods \citep{r.2.2.3,r.2.2.4,r.2.2.5,r.2.2.7} simulate the effect of Bayesian neural networks by performing multiple forward propagation in the inference stage, and statistical properties of different predictions (such as mean and variance) are used to estimate the uncertainty of the model. However, multiple forward propagation leads to high computational cost. \citet{chen2022sample,chen2023generalized,2024chenOE2,chen2024reduced} and \citet{2024chenOE} combine model order reduction techniques with deep learning to expedite the sampling process for uncertainty quantification. Compared with aforementioned uncertainty quantification methods, the Laplace approximation \citep{r.2.2.12,r.2.2.13} does not require retraining the model or modifying the training process. Instead, it approximates the posterior distribution with a Gaussian distribution around the mode of the posterior, thereby effectively reducing computational costs. Specifically, by performing a Taylor expansion of the log-posterior and retaining terms up to the second order at the mode, a Gaussian distribution is obtained to approximate the posterior. The mean of this Gaussian distribution is the mode, and the covariance is the negative inverse of the Hessian matrix at the mode.

\subsection{Uncertainty in NeRFs}
In practical applications, NeRFs needs to address uncertainty issues to improve model reliability and robustness. Therefore, researchers have explored various approaches. ActiveNeRF \citep{activenerf} models the radiance values at each location as a Gaussian distribution instead of a single value, enabling NeRFs to provide reasonable high-variance predictions in unobserved regions. S-NeRF \citep{s-nerf} utilizes Bayesian variational inference to achieve uncertainty quantification related to outputs in tasks such as novel view synthesis or depth estimation. CF-NeRF \citep{cf-nerf} employs conditional normalizing flows and latent variable modelling to flexibly learn the radiance field distribution without any prior assumptions. It estimates uncertainty by evaluating the predicted mean and variance during the inference process. \citet{density-aware} proposes to quantify prediction uncertainty by independently training multiple NeRFs with different initial parameters on the same dataset \citep{lxh,wjq}. In addition to considering the variance in RGB colors, it introduces an epistemic uncertainty term based on termination probability to capture uncertainty in unobserved regions. FG-NeRF \citep{fg-nerf} decouples NeRFs into deterministic and probabilistic branches, using Flow-GAN \citep{Flow-GAN} to model the probabilistic branch to avoid independence assumptions. Additionally, it employs a patch adversarial training strategy. These designs enable it to achieve more accurate uncertainty estimation in complex scenes. ProbNeRF \citep{probnerf} employs the hamiltonian monte carlo (HMC) \citep{HMC} method during testing to perform posterior inference on NeRFs' parameters for given views. It accurately infers the 3D geometry and appearance of objects from a single or several views while quantifying the associated uncertainty. Recursive-NeRF \citep{recursive-nerf} draws inspiration from level of detail (LOD). Each network layer predicts the uncertainty of query points: for points with low uncertainty, results are directly output without passing to deeper layers; for points with high uncertainty, they are passed to the next, more powerful layer for further processing. This strategy significantly improves rendering efficiency while ensuring the quality of view synthesis. Unlike the above methods, we introduce an additional perturbation field to avoid retraining the model or modifying the training process, significantly reducing the computational cost.

\begin{figure}
	\centering
	\includegraphics[width=1.0\textwidth]{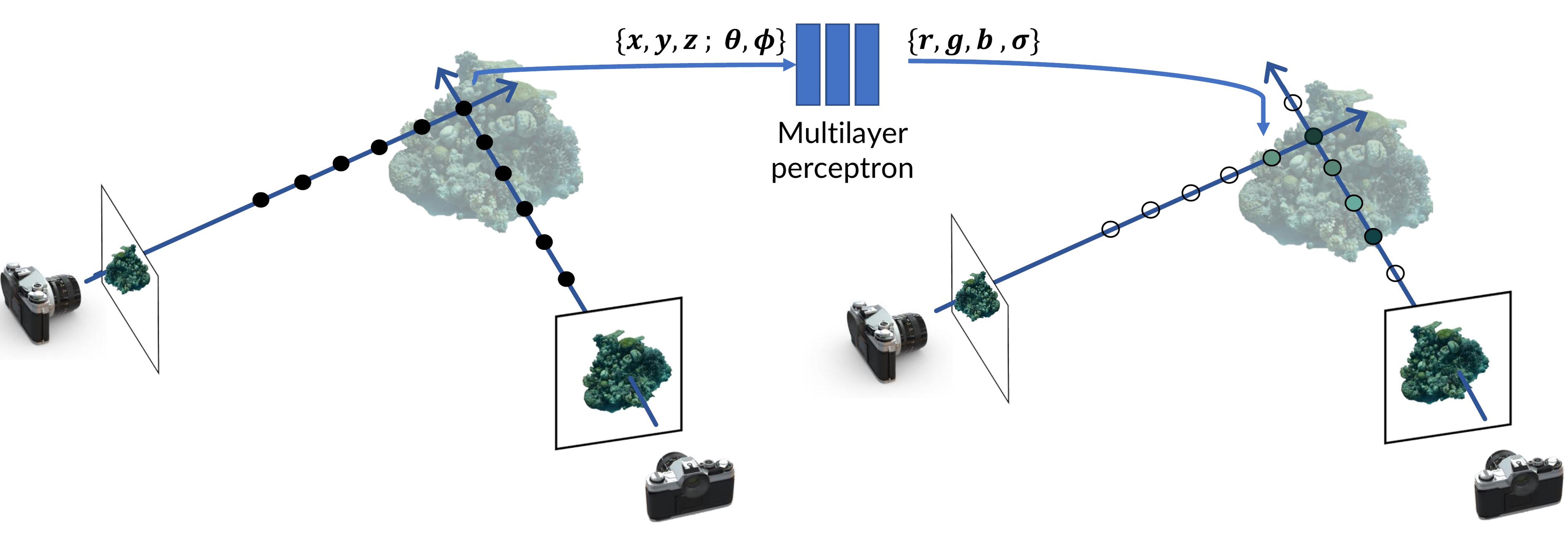}
	\caption{NeRFs architecture \citep{seathrunerf}. NeRFs utilizes a MLP network, taking each 3D spatial coordinate $\mathbf{x}=(x,y,z)$ and viewing direction $\mathbf{d}=(\theta,\phi)$ in the scene as input, and outputs the corresponding color $\mathbf{c}=(r,g,b)$ and density $\sigma$.}
	\label{FIG:1}
\end{figure}

\section{Scientific background}
\subsection{SeaThru}

Traditional underwater image generation models based on atmospheric models result in significant errors. To address this issue of underwater image color restoration, SeaThru \citep{Sea-thru} employs a revised optical model that accounts for the attenuation and scattering characteristics of light propagation in water, thereby generating images that are closer to true colors.

The underwater image generation model is described as follows:
\begin{equation}I_\mathbf{c}=D_\mathbf{c}+B_\mathbf{c}\label{Eq.1}\end{equation}
where $\mathbf{c}{\in}\{RGB\}$ is color channel, $I_\mathbf{c}$ is the image captured by the camera, $D_\mathbf{c}$ is the direct signal, and $B_\mathbf{c}$ is the backscatter.

Traditional methods assume a uniform attenuation coefficient for light throughout the entire scene, which is a coarse approximation. Actually, the direct signal and backscatter are controlled by different attenuation coefficients, which also depend on factors such as object distance and reflectance. Therefore, SeaThru extends Eq. (\ref{Eq.1}) as follows:
\begin{equation}
I_\mathbf{c}=\overbrace{J_{\mathbf{c}}\cdot\underbrace{(e^{-\beta^D(\mathbf{v}_D)\cdot z})}_{\text{attenuation}}}^{\text{direct}}+\overbrace{B_{\mathbf{c}}^{\infty}\cdot\underbrace{(1-e^{-\beta^B(\mathbf{v}_B)\cdot z})}_{\text{attenuation}}}^{\text{backscatter}}
\end{equation}
where $J_{\mathbf{c}}$ is the clear scene that should be captured ideally, $B_{\mathbf{c}}^\infty$ represents veiling light, $z$ is the distance from the object to the camera, $\beta^{D}$ and $\beta^{B}$ are the attenuation coefficients and backscatter coefficients, respectively. The vectors $\mathbf{v}_{D}$ and $\mathbf{v}_{B}$ represent the dependence of $\beta^{D}$ and $\beta^{B}$ on factors such as distance, reflectance.

SeaThru analyzes multi-view images to estimate the distance of each pixel, thereby inferring the effects of scattering and attenuation. It then corrects color distortions in the image by applying these estimated values.

\subsection{NeRFs}

NeRFs is a novel 3D reconstruction technique that reconstructs high-quality 3D scenes from collections of 2D images and camera poses. It achieves high-resolution rendering and image synthesis by representing scenes as a dense volumetric radiance field function and leveraging volumetric rendering techniques.

Specifically, as shown in the Fig. \ref{FIG:1}, NeRFs utilizes MLP to learn a mapping function $F$ that maps any given 3D spatial point $\mathbf{x}=(x,y,z)$ and viewing direction $\mathbf{d}=(\theta,\phi)$ to a color value $\mathbf{c}=(r,g,b)$ and a density value $\sigma$:
\begin{equation}F_{\boldsymbol{\varphi}}(\mathbf{x},\mathbf{d})=(\mathbf{c},\sigma)\end{equation}
where $\boldsymbol{\varphi}$ are the learnable parameters of the MLP.

To render the continuous 5D scene structure represented by color $\mathbf{c}$ and density $\sigma$ into a 2D image, NeRFs uses the volumetric rendering equation to integrate the color and density values along a camera ray $\mathbf{r}(t)=\mathbf{o}+t\mathbf{d}$, where $\mathbf{o}$ is the camera center and $t\in\mathbb{R}_{+}$. 

The expected color obtained along the ray $\mathbf{r}$ is given by:
\begin{equation}
C(\mathbf{r})=\int_{t_n}^{t_f}T(t)\sigma(t)\mathbf{c}(t)dt\label{Eq.4}
\end{equation}
Here, the integration is bounded between the near bound $t_{n}$ and the far bound $t_{f}$, and the accumulated transmittance $T(t)$ between them is defined as:
\begin{equation}
T(t)=\exp\biggl(-\int_{t_{n}}^{t}\sigma(s)ds\biggr)\label{Eq.5}\end{equation}

In practice, NeRFs utilizes the quadrature rule \citep{quadrature_rule} to discretize the integration range $[t_n,t_f]$ into $N$ intervals $\{[t_i,t_{i+1}]\}_{i=1}^{N}$ (where $t_n=t_1<\cdots<t_N=t_f$), and assumes that the density $\sigma$ and color $\mathbf{c}$ are constant within each interval. Thus:
\begin{equation}\hat{C}(\mathbf{r})=\sum_{i=1}^N\int_{t_i}^{t_{i+1}}T(t)\sigma_i\mathbf{c}_idt=\sum_{i=1}^NT(t_i)(1-\exp(-\sigma_i\delta_i))\mathbf{c}_i,\end{equation}
and
\begin{equation}T(t_{i})=\exp\left(-\sum_{j=0}^{i-1}\sigma_{j}\delta_{j}\right)\end{equation}
where $\delta_{i}=t_{i+1}-t_{i}$ is the distance between adjacent sample points.

NeRFs adjusts the parameters of the MLP during training by optimizing the squared distance between the expected color $\hat{C}(\mathbf{r})$ and the ground truth $C^{\mathrm{gt}}(\mathbf{r})$.

\subsection{SeaThru-NeRF}\label{section:SeaThru-NeRF}

\begin{figure}%
	\centering
	\includegraphics[width=.6\columnwidth]{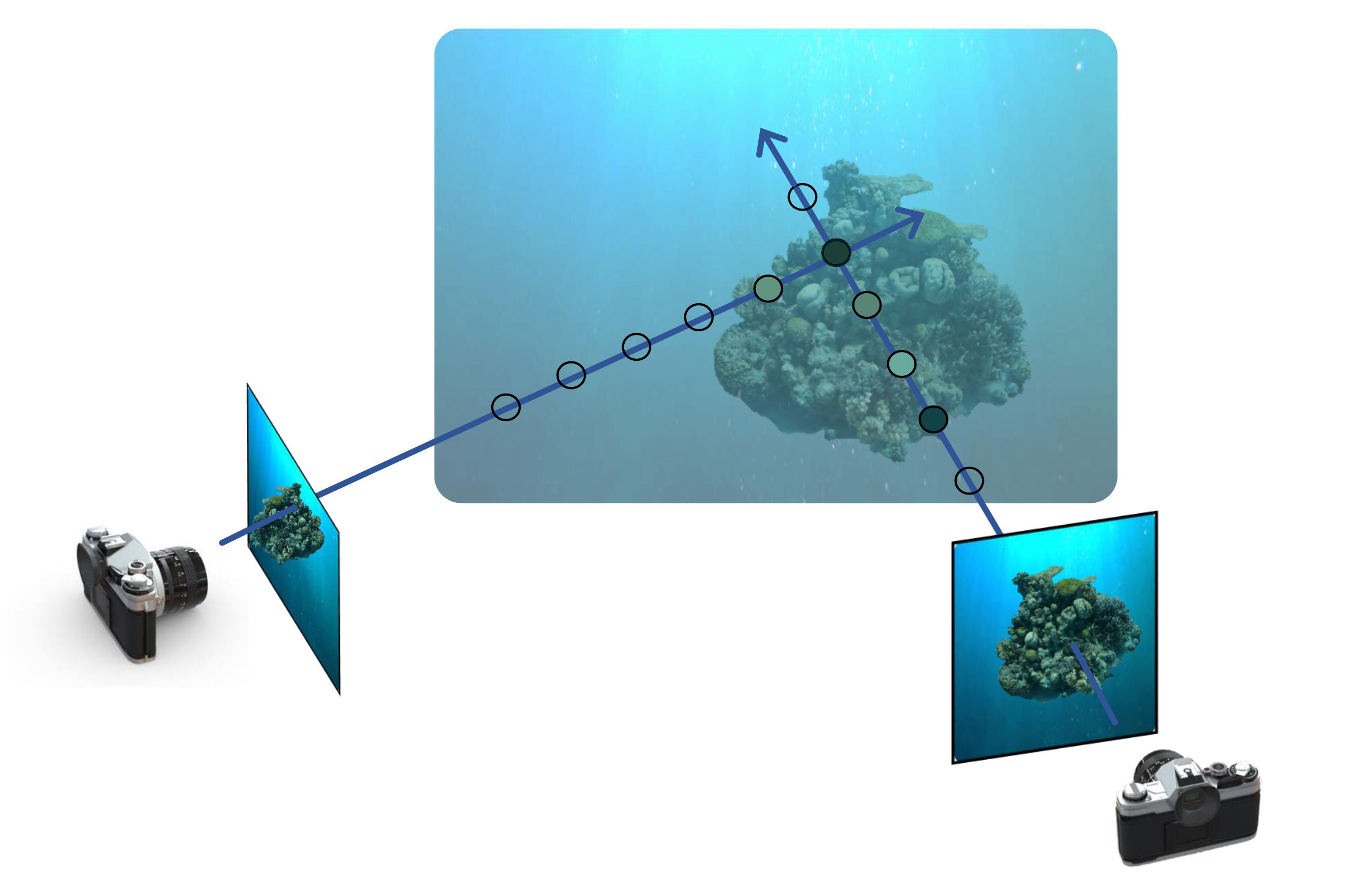}
      \caption{SeaThru-NeRF application scenario \citep{seathrunerf}.}
      \label{FIG:2}
\end{figure}

Inspired by SeaThru \citep{Sea-thru}, SeaThru-NeRF not only considers the opaque objects in traditional NeRFs but also treats the medium as a semi-transparent entity, introducing separate color and density parameters for both the medium and the objects. This method extends the capability of NeRFs to handle scattering media such as underwater (Fig. \ref{FIG:2}) through mapping function $F$.

Specifically, it replaces the traditional single opaque object density  $\sigma(t)$ with the sum of the object density $\sigma^{\mathrm{obj}}(t)$ and the medium density $\sigma^{\mathrm{med}}(t)$. 
\begin{equation}C(\mathbf{r})=\int_{t_n}^{t_f}T(t)\Big(\sigma^{\text{obj}}(t)\mathbf{c}^{\text{obj}}(t)+\sigma^{\text{med}}(t)\mathbf{c}^{\text{med}}(t)\Big)dt\end{equation}

\begin{equation}T(t)=\exp\biggl(-\int_{t_n}^t\bigl(\sigma^{\mathrm{obj}}(s)+\sigma^{\mathrm{med}}(s)\bigr)ds\biggr)\end{equation}
where $\mathbf{c}^{\mathrm{obj}}(t)$ and $\mathbf{c}^{\mathrm{med}}(t)$ are the color of the object and medium at $t$, respectively. When $\sigma^{\mathrm{med}}=0$, it simplifies to the traditional NeRFs case.

Adopting the same discretization strategy as NeRFs:
\begin{equation}
\hat{C}(\mathbf{r})=\sum_{i=1}^NT(t_i)\biggl(1-\exp({-(\sigma_i^{\mathrm{obj}}+\sigma_i^{\mathrm{obj}})\delta_i})\biggr)\frac{\sigma_i^{\mathrm{obj}}\mathbf{c}_i^{\mathrm{obj}}+\sigma_i^{\mathrm{med}}\mathbf{c}_i^{\mathrm{med}}}{\sigma_i^{\mathrm{obj}}+\sigma_i^{\mathrm{med}}}\label{Eq.11}
\end{equation}

\begin{equation}
T(t_i)=\exp\Bigg(-\sum_{j=0}^{i-1}(\sigma_j^{\text{obj}}+\sigma_j^{\text{med}})\delta_j\Bigg)
\end{equation}

According to the rendering equation Eq. (\ref{Eq.11}), the color contributions can be divided into object and medium components, reflecting their different contributions to the final color.
\begin{equation}\hat{C}(\mathbf{r})=\sum_{i=1}^N\hat{C}_i^{\mathrm{obj}}(\mathbf{r})+\sum_{i=1}^N\hat{C}_i^{\mathrm{med}}(\mathbf{r})\end{equation}
where
\begin{equation}\hat{C}_{i}^{\mathrm{obj}}(\mathbf{r})=T(t_{i})\biggl(1-\exp({-(\sigma_{i}^{\mathrm{obj}}+\sigma_{i}^{\mathrm{med}})\delta_{i}})\biggr)\frac{\sigma_{i}^{\mathrm{obj}}\mathbf{c}_{i}^{\mathrm{obj}}}{\sigma_{i}^{\mathrm{obj}}+\sigma_{i}^{\mathrm{med}}}\end{equation}
and
\begin{equation}\hat{C}_{i}^{\text{med}}(\mathbf{r})=T(t_{i})\biggl(1-\exp({-(\sigma_{i}^{\text{obj}}+\sigma_{i}^{\text{med}}) \delta_{i}})\biggr)\frac{\sigma_{i}^{\text{med}}\mathbf{c}_{i}^{\text{med}}}{\sigma_{i}^{\text{obj}}+\sigma_{i}^{\text{med}}}\end{equation}

To simplify the model, SeaThru-NeRF assumes the color and density of the medium are constant along the ray $\mathbf{r}$. Additionally, because of $\sigma^{\mathrm{med}}\gg\sigma^{\mathrm{obj}}$ before the object and $\mathbf{\sigma}^{\mathrm{med}}\ll\sigma^{\mathrm{obj}}$ at the object \citep{seathrunerf}, thus

\begin{equation}\hat{C}_i^{\text{obj}}(\mathbf{r})=T_i\cdot\left(1-\exp({-\sigma_i^{\text{obj}}\delta_i})\right)\cdot\mathbf{c}_i^{\text{obj}}\end{equation}

\begin{equation}\hat{C}_i^{\text{med}}(\mathbf{r})=T_i\cdot\left(1-\exp({-\sigma^{\text{med}}\delta_i})\right)\cdot\mathbf{c}^{\text{med}}\end{equation}

\begin{equation}T_i=\exp\left(-\sum_{j=0}^{i-1}\sigma_j^{\text{obj}}\delta_j\right)\cdot\exp\left(-\mathbf{\sigma}^{\text{med}}t_i\right)\end{equation}

In the aforementioned discussion, the object component and the backscatter component used the same attenuation coefficient. According to SeaThru, the effective $\sigma^{\mathrm{med}}$ for the object component $\hat{C}^{\mathrm{obj}}(\mathbf{r})$ and the backscatter component $\hat{C}^{\mathrm{med}}(\mathbf{r})$ are different. Therefore, in the final model, different parameters are used for each component—$\sigma^{\mathrm{attn}}$ for the object component $\hat{C}^{\mathrm{obj}}(\mathbf{r})$ and $\sigma^{\mathrm{bs}}$ for the backscatter component $\hat{C}^{\mathrm{med}}(\mathbf{r})$.

\begin{equation}\hat{C}_{i}^{\mathrm{obj}}(\mathbf{r})=T_{i}^{\mathrm{obj}}\cdot\exp\Bigl(-\sigma^{\mathrm{attn}}t_{i}\Bigr)\cdot\Bigl(1-\exp(-\sigma_{i}^{\mathrm{obj}}\delta_{i})\Bigr)\cdot\mathbf{c}_{i}^{\mathrm{obj}}\end{equation}

\begin{equation}\hat{C}_i^{\mathrm{med}}(\mathbf{r})=T_i^{\mathrm{obj}}\cdot\exp\Bigl(-\mathbf{\sigma}^{\mathrm{bs}}t_i\Bigr)\cdot\Bigl(1-\exp(-\mathbf{\sigma}^{\mathrm{bs}}\delta_i)\Bigr)\cdot\mathbf{c}^{\mathrm{med}}\end{equation}

\begin{equation}T_{i}^{\text{obj}}=\exp\Bigg(-\sum_{j=0}^{i-1}\sigma_{j}^{\text{obj}}\delta_{j}\Bigg)\end{equation}

Like NeRFs, SeaThru-NeRF through minimize the squared distance between expected color and ground truth to optimize the network parameters for
each ray $\mathbf{r}$ sampled from image $\mathbf{I}_{n}$ of training set images $\mathcal{I}=\{\mathbf{I}\}_{n=0}^{\mathrm{N}}$. In a Bayesian perspective, this is equivalent to assuming a Gaussian likelihood $p(C_{\boldsymbol{\varphi}}|\boldsymbol{\varphi})\sim\mathcal{N}(C_{n}^{\mathrm{gt}},\frac12)$ and inferring $\boldsymbol{\varphi}^{*}$, the mode of the posterior distribution
\begin{equation}\boldsymbol{\varphi}^{*}=\arg\max_{\boldsymbol{\varphi}}p(\boldsymbol{\varphi}|\mathcal{I})\end{equation}
which, by Bayes’ rule, is the same as minimizing the negative log-likelihood
\begin{equation}{\boldsymbol{\varphi}}^*=\arg\min_{\boldsymbol{\varphi}} \mathbb{E}_i \mathbb{E}_{\mathbf{r}\sim\mathbf{I}_n}\|\hat{C}_{\boldsymbol{\varphi}}(\mathbf{r})-C_n^{\mathrm{gt}}(\mathbf{r})\|_2^2\end{equation}

\section{Uncertainty estimation}
\subsection{Neural Laplace approximations}
Laplace approximation obtains the optimal network weights $\boldsymbol{\omega}^{*}$ through pre-trained model, then approximates the posterior distribution of the network parameters as a multivariate Gaussian distribution centered at $\boldsymbol{\omega}^{*}$, i.e.,
\begin{equation}p(\boldsymbol{\omega}|\mathcal{I})\sim\mathcal{N}(\boldsymbol{\omega}^*,\Sigma)\end{equation}
where $\mathrm{\Sigma}$ is the covariance matrix. 

According to the Laplace approximation, we consider the second-order Taylor expansion of the objective function $h(\boldsymbol{\omega})=-\log p(\boldsymbol{\omega}|\mathcal{I})$ at $\boldsymbol{\omega}^{*}$:
\begin{equation}h(\boldsymbol{\omega})\approx h(\boldsymbol{\omega}^*)+\frac{1}{2}(\boldsymbol{\omega}-\boldsymbol{\omega}^*)^T\mathbf{H}(\boldsymbol{\omega}^*)(\boldsymbol{\omega}-\boldsymbol{\omega}^*)\label{Eq.24}\end{equation}
where $H(\boldsymbol{\omega}^*)$ is the Hessian matrix of $h(\boldsymbol{\omega})$ at $\boldsymbol{\omega}^{*}$, and since $\boldsymbol{\omega}^{*}$ is an extremum point of $h(\boldsymbol{\omega})$ , the first-order derivative term is zero.

By comparing the Eq. (\ref{Eq.24}) with the usual log squared exponential Gaussian likelihood of the multivariate Gaussian distribution, we can derive the expression for the covariance matrix $\Sigma$:
\begin{equation}\Sigma=-\mathbf{H}(\boldsymbol{\omega}^*)^{-1}\end{equation}

However, directly equating $\boldsymbol{\omega}$ with $\boldsymbol{\varphi}$ is impracticable. The high correlation between model parameters makes accurately estimating the covariance matrix $\Sigma$ of the parameter distribution challenging. Additionally, even with an accurate $\Sigma$, converting it into geometrically meaningful distribution necessitates an expensive sampling process, resulting in significant computational overhead.

To address these issues, we introduce a reparameterization method based on a perturbation field in Section \ref{senction4.2}, which is equivalent to adding a differentiable spatial deformation module before the MLP. This reparameterization method makes the parameters more amenable to Laplace approximation.

\begin{figure}
	\centering
	\begin{subfigure}{0.325\linewidth}
		\centering
		\includegraphics[width=0.8\linewidth]{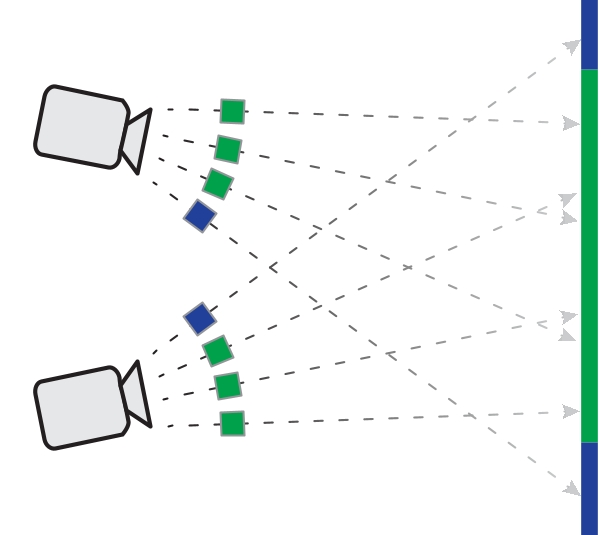}
            \caption{}
		\label{3a}
	\end{subfigure}
	\centering
	\begin{subfigure}{0.325\linewidth}
		\centering
		\includegraphics[width=0.8\linewidth]{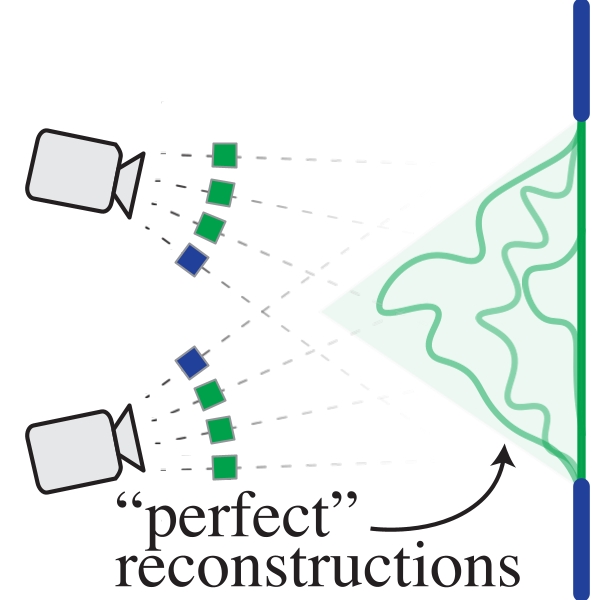}
            \caption{}
		\label{3b}
	\end{subfigure}
	\centering
	\begin{subfigure}{0.325\linewidth}
		\centering
		\includegraphics[width=0.8\linewidth]{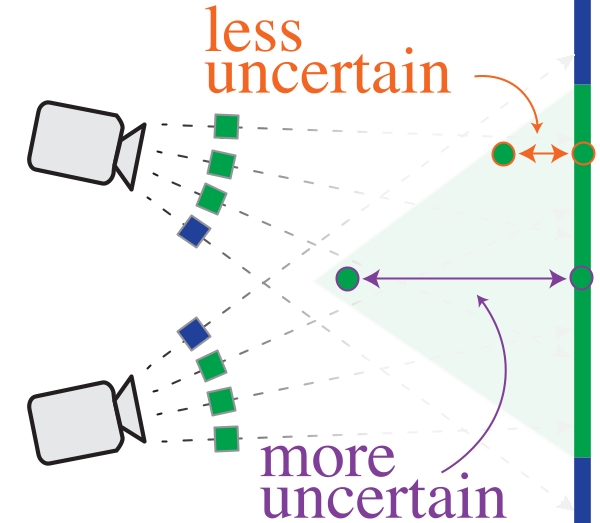}
            \caption{}
		\label{3c}
	\end{subfigure}
	\caption{\citet{bayes_rays} considers a two-dimensional plane with a solid blue line segment and a green center line segment. Assuming two cameras capture the scene within a 60-degree cone, it can be observed that the green line segment can be replaced by many possible curves, all of which can achieve 'perfect' photometric reconstruction based on the captured pixels.}
	\label{FIG:3}
\end{figure}

\subsection{Modeling perturbations}\label{senction4.2}

As illustrated in Fig. \ref{FIG:3}, a space (green area) exists where the green line segment in Fig. \ref{3a} can be perturbed to any shape within this region without impacting the reconstruction loss (Fig. \ref{3b}). During model training, different random seeds may cause convergence to various configurations within this space. Consequently, for a pre-trained reconstruction model, due to the limited training data, certain regions of the scene have perturbable spaces that do not affect the reconstruction loss. The degree of allowable perturbation indicates the model's uncertainty level in that region. By systematically applying perturbations throughout the entire 3D scene and measuring the maximum tolerated perturbation at each spatial position, one can obtain the model's uncertainty distribution across the entire space (Fig. \ref{3c}).

\begin{figure}
	\centering
	\includegraphics[width=0.7\textwidth]{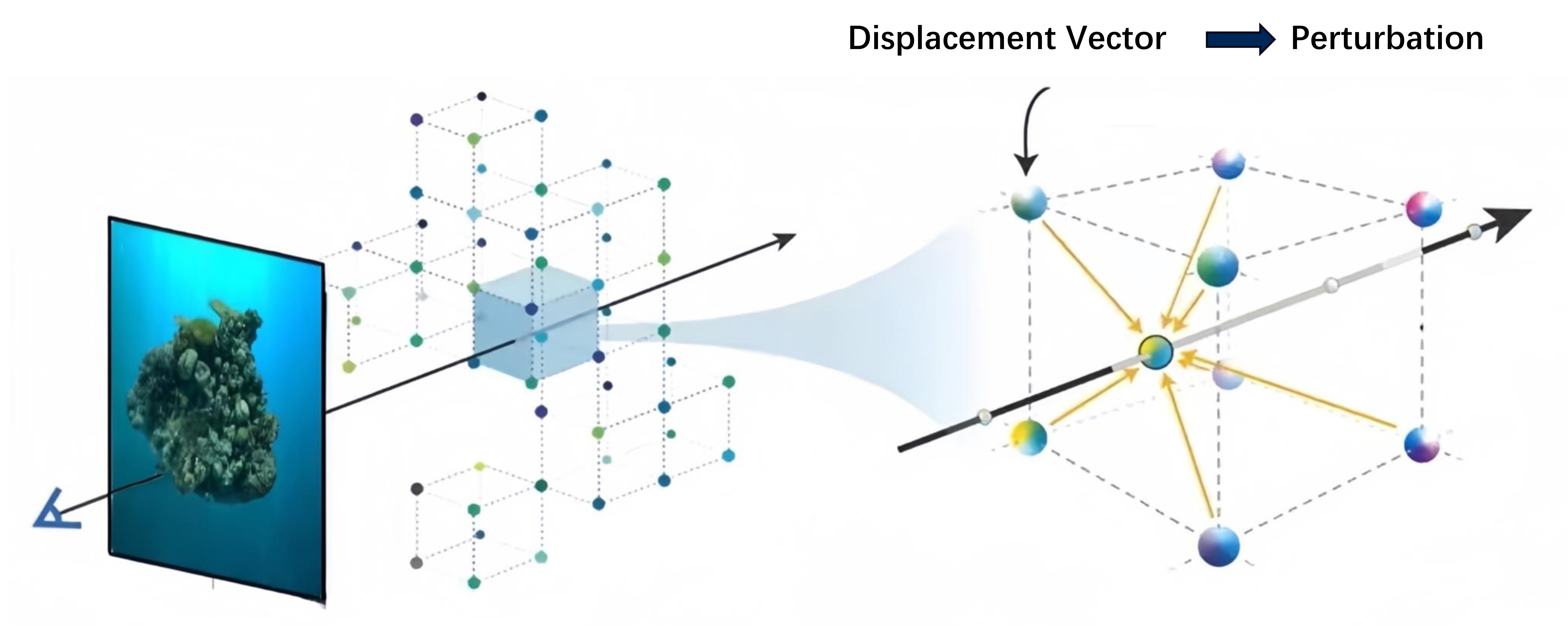}
	\caption{Perturbation.}
	\label{FIG:perturbation}
\end{figure}

Inspired by the above, we introduce a parametrized perturbation field $\mathcal{D}{:}\mathbb{R}^{D}\to\mathbb{R}^{D}$, which can be interpreted as a perturbation transformation applied to the coordinates before inputting them into the MLP (Fig. \ref{FIG:perturbation}). The parameters of the perturbation field $\mathcal{D}$ are represented as $\boldsymbol{\omega}\in\mathbb{R}^{M^{D}\times D}$ , where $M$ denotes the size of the grid used to store displacement vectors and $D$ represents vector dimension. For any spatial coordinate $\mathbf{x}$, its perturbation is computed by trilinear interpolation of displacement vectors from neighboring grid points:
\begin{equation}\mathcal{D}_{\boldsymbol{\omega}}(\mathbf{x})=\mathrm{Trilinear}(\mathbf{x},\boldsymbol{\omega})\end{equation}

Next, we reparameterize the optimized MLP neural network by perturbing each coordinate:

\begin{equation}\tilde{\sigma}_{\boldsymbol{\omega}}^{\mathrm{obj}}(\mathbf{x})=\sigma_{{\boldsymbol{\varphi}}^{*}}^{\mathrm{obj}}(\mathbf{x}+\mathcal{D}_{\boldsymbol{\omega}}(\mathbf{x}))\end{equation}

\begin{equation}\tilde{\mathbf{c}}_{\boldsymbol{\omega}}^{\mathrm{obj}}(\mathbf{x})=\mathbf{c}_{{\boldsymbol{\varphi}}^{*}}^{\mathrm{obj}}(\mathbf{x}+\mathcal{D}_{\boldsymbol{\omega}}(\mathbf{x}),\mathbf{d})\end{equation}
where $\sigma_{{\boldsymbol{\varphi}}^{*}}^{\mathrm{obj}}$ and $\mathbf{c}_{{\boldsymbol{\varphi}}^{*}}^{\mathrm{obj}}$ are the optimized density and radiance of object, respectively. And as mentioned in Section \ref{section:SeaThru-NeRF}, the density and color associated with the medium are constants at each spatial coordinate, so they are not affected by perturbations.

Based on the reparameterized $\tilde{\sigma}_{\boldsymbol{\omega}}^{\mathrm{obj}}$ and $\tilde{\mathbf{c}}_{\boldsymbol{\omega}}^{\mathrm{obj}}$ , the perturbed predicted pixel color can be obtained as:
\begin{equation}\tilde{C}_{\boldsymbol{\omega}}(\mathbf{r})=\sum_{i=1}^N\tilde{C}_i^{\mathrm{obj}}(\mathbf{r})+\sum_{i=1}^N\tilde{C}_i^{\mathrm{med}}(\mathbf{r})\end{equation}
where
\begin{equation}\tilde{C}_{i}^{\mathrm{obj}}(\mathbf{r})=\tilde{T}_{i}^{\mathrm{obj}}\cdot\exp\Bigl(-\sigma^{\mathrm{attn}}t_{i}\Bigr)\cdot\Bigl(1-\exp(-\tilde{\sigma}_{i}^{\mathrm{obj}}\delta_{i})\Bigr)\cdot\tilde{\mathbf{c}}_{i}^{\mathrm{obj}}\end{equation}
\begin{equation}\tilde{C}_i^{\mathrm{med}}(\mathbf{r})=\tilde{T}_i^{\mathrm{obj}}\cdot\exp\Bigl(-\mathbf{\sigma}^{\mathrm{bs}}t_i\Bigr)\cdot\Bigl(1-\exp(-\mathbf{\sigma}^{\mathrm{bs}}\delta_i)\Bigr)\cdot\mathbf{c}^{\mathrm{med}}\end{equation}
and accumulated transmittance of object can be obtained as:
\begin{equation}\tilde{T}_{i}^{\text{obj}}=\exp\Bigg(-\sum_{j=0}^{i-1}\tilde{\sigma}_{j}^{\text{obj}}\delta_{j}\Bigg)\end{equation}

Since we still want the predicted color ${\tilde{C}}_{\boldsymbol{\omega}}(\mathbf{r})$ to be as close to $\mathrm{C}_{n}^{\mathrm{gt}}$ as possible after reparameterization, we assume a likelihood function of the same form as the original model, i.e., $\tilde{C}_{\boldsymbol{\omega}}|{\boldsymbol{\omega}}\sim\mathcal{N}(C_{n}^{\mathrm{gt}},\frac{1}{2})$. Moreover, since ${\boldsymbol{\varphi}}^{*}$ are the optimal parameters obtained during the training of the MLP network, the model has achieved optimal performance at these parameters. Therefore, when perturbations are applied to new parameters, we expect these perturbations to neither significantly improve nor degrade the model's performance. Hence, we impose a regularization Gaussian prior $\boldsymbol{\omega}\sim\mathcal{N}(0,\lambda^{-1})$ on the new parameters $\boldsymbol{\omega}$.

Under these assumptions, the negative log-likelihood function of the posterior distribution $p(\boldsymbol{\omega}|\mathcal{I})$ is given by:
\begin{equation}h(\boldsymbol{\omega})=\mathbb{E}_n\mathbb{E}_{\mathbf{r}\sim\mathbf{I}_n}\|\tilde{C}_{\boldsymbol{\omega}}(\mathbf{r})-C_n^{\mathrm{gt}}(\mathbf{r})\|_2^2+\lambda\| \boldsymbol{\omega}\|^2\end{equation}
Here, $\mathbb{E}_n\mathbb{E}_{\mathbf{r}\sim\mathbf{I}_n}\|\tilde{C}_{\boldsymbol{\omega}}(\mathbf{r})-C_n^{\mathrm{gt}}(\mathbf{r})\|_2^2$ represents the reconstruction error of the pixel colors, and $\lambda\| \boldsymbol{\omega}\|^2$ is the regularization term.

When $\boldsymbol{\omega}=0$, we have $\tilde{\sigma}_{0}^{\mathrm{obj}}(\mathbf{x})=\sigma_{{\boldsymbol{\varphi}}^{*}}^{\mathrm{obj}}(\mathbf{x})$ and $\tilde{\mathbf{c}}_{0}^{\mathrm{obj}}(\mathbf{x})=\mathbf{c}_{{\boldsymbol{\varphi}}^{*}}^{\mathrm{obj}}(\mathbf{x},\mathbf{d})$, thus $\tilde{C}_{0}(\mathbf{r})=C_{{\boldsymbol{\varphi}}^{*}}(\mathbf{r})$. Therefore, $\boldsymbol{\omega}=0$ minimizes the reconstruction error and is the mode of the posterior distribution $p(\boldsymbol{\omega}|\mathcal{I})$.

According to the Laplace approximation, we perform a second-order Taylor expansion around the mode, which yields
\begin{equation}\Sigma=-\mathbf{H}(0)^{-1}\end{equation}
where $\mathbf{H}(0)$ is the Hessian matrix of $h(\boldsymbol{\omega})$ evaluated at zero.

\subsection{Approximating H}

Because directly computing the second derivatives in the Hessian matrix is computationally expensive, we approximate it using the Fisher information.

For any parameterized family of probability distributions $p_{\boldsymbol{\omega}}$ , the Hessian matrix of its log-likelihood function with respect to the parameters ${\boldsymbol{\omega}}$ is related to the Fisher information as follows:
\begin{equation}\mathcal{I}(\boldsymbol{\omega})=-\mathbb{E}_{\mathbf{X}\sim p_{\boldsymbol{\omega}}}\left[\frac{\partial^2h(\mathbf{X};\boldsymbol{\omega})}{\partial\boldsymbol{\omega}^2}\Big| \boldsymbol{\omega}\right]=-\mathbf{H}(\boldsymbol{\omega})\label{Eq.34}\end{equation}
where $h(\mathbf{X};\boldsymbol{\omega})$ is defined as the log-likelihood function.

Furthermore, under reasonable regularity conditions, the Fisher information can also be defined as:
\begin{equation}\mathcal{I}(\boldsymbol{\omega})=\mathbb{E}_{\mathbf{X}\sim p_{\boldsymbol{\omega}}}\left[\frac{\partial h(\mathbf{X};\boldsymbol{\omega})}{\partial\boldsymbol{\omega}}^{\top}\frac{\partial h(\mathbf{X};\boldsymbol{\omega})}{\partial\boldsymbol{\omega}}\big|\boldsymbol{\omega}\right]\label{Eq.35}\end{equation}

We use the random variable $(\mathbf{r},\mathbf{y})$ to correspond to a camera ray $\mathbf{r}$ and its corresponding ground truth $\mathbf{y}=C_n^{\mathrm{gt}}(\mathbf{r})$. Thus,
\begin{equation}\mathcal{I}(\boldsymbol{\omega})=\mathbb{E}_{(\mathbf{r},\mathbf{y})}\Big[4\epsilon_{\boldsymbol{\omega}}(\mathbf{r})\mathbf{J}_{\boldsymbol{\omega}}(\mathbf{r})^{\top}\mathbf{J}_{\boldsymbol{\omega}}(\mathbf{r})\Big]+2\lambda\mathbf{I}\end{equation}
where $\epsilon_{\boldsymbol{\omega}}(\mathbf{r})=\|\tilde{C}_{\boldsymbol{\omega}}(\mathbf{r})-C_{n}^{\mathrm{gt}}(\mathbf{r})\|^2$. And
\begin{equation}\mathbf{J}_{\boldsymbol{\omega}}(\mathbf{r})=\frac{\partial\tilde{C}_{\boldsymbol{\omega}}(\mathbf{r})}{\partial\boldsymbol{\omega}}\end{equation}
represents the Jacobian matrix of the predicted color with respect to the parameter $\boldsymbol{\omega}$, which can be computed through backpropagation.

Furthermore, based on the properties of conditional expectations:
\begin{equation}\mathcal{I}(\boldsymbol{\omega})=\mathbb{E}_{\mathbf{r}}\Big[4\mathbb{E}_{\mathbf{y}|\mathbf{r}}\left[\epsilon_{\boldsymbol{\omega}}(\mathbf{r})\right]\mathbf{J}_{\boldsymbol{\omega}}(\mathbf{r})^{\top}\mathbf{J}_{\boldsymbol{\omega}}(\mathbf{r})\Big]+2\lambda\mathbf{I}\end{equation}

According to the likelihood ${\mathcal{N}}(C_{n}^{\mathrm{gt}},{\frac{1}{2}})$, it holds that $\mathbb{E}_{\mathbf{y}|\mathbf{r}}\left[\epsilon_{\boldsymbol{\omega}}(\mathbf{r})\right]$ is nothing more than $\frac{1}{2}$, so:
\begin{equation}\mathcal{I}(\boldsymbol{\omega})=\mathbb{E}_{\mathbf{r}}\Big[2\mathbf{J}_{\boldsymbol{\omega}}(\mathbf{r})^\top\mathbf{J}_{\boldsymbol{\omega}}(\mathbf{r})\Big]+2\lambda\mathbf{I}\end{equation}

Here, we approximate the expectation by sampling $R$ rays.
\begin{equation}\mathbb{E}_{\mathbf{r}}\Big[\mathbf{J}_{\boldsymbol{\omega}}(\mathbf{r})^\top\mathbf{J}_{\boldsymbol{\omega}}(\mathbf{r})\Big]\approx\frac1R\sum_{\mathbf{r}}\mathbf{J}_{\boldsymbol{\omega}}(\mathbf{r})^\top\mathbf{J}_{\boldsymbol{\omega}}(\mathbf{r})\end{equation}

We can obtain:
\begin{equation}\mathcal{I}(\boldsymbol{\omega})\approx\frac2R\sum_{\mathbf{r}}\mathbf{J}_{\boldsymbol{\omega}}(\mathbf{r})^\top\mathbf{J}_{\boldsymbol{\omega}}(\mathbf{r})+2\lambda\mathbf{I}\end{equation}

From Eq. (\ref{Eq.34}), we can finally obtain:
\begin{equation}\mathbf{H}(\boldsymbol{\omega})\approx-\frac{2}{R}\sum_{\mathbf{r}}\mathbf{J}_{\boldsymbol{\omega}}(\mathbf{r})^\top\mathbf{J}_{\boldsymbol{\omega}}(\mathbf{r})-2\lambda\mathbf{I}\end{equation}

\begin{figure}
	\centering
	\includegraphics[width=0.7\textwidth]{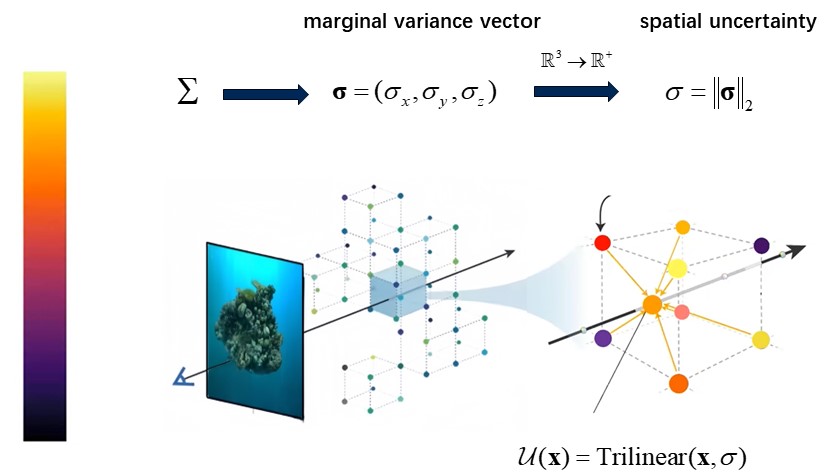}
	\caption{Spatial uncertainty.}
	\label{FIG:spatial_uncertainty}
\end{figure}

\subsection{Spatial uncertainty}

Since each entry of the parameter vector $\boldsymbol{\omega}$ corresponds to a vertex in the grid, its influence is limited to the cell containing that vertex, making $\mathbf{H}(\boldsymbol{\omega})$ inherently sparse and thereby minimizing the number of related parameters. Similar to the approach by \citet{ritter2018a}, we approximate $\Sigma$ by considering only the diagonal elements of $\mathbf{H}$.
\begin{equation}\Sigma\approx\mathrm{diag}\Bigg(\frac{2}{R}\sum_{\mathbf{r}}\mathbf{J}_{\boldsymbol{\omega}}(\mathbf{r})^{\top}\mathbf{J}_{\boldsymbol{\omega}}(\mathbf{r})+2\lambda\mathbf{I}\Bigg)^{-1}\end{equation}
where $\Sigma$ encodes the spatial uncertainty of the radiance field. Intuitively, it represents the extent to which the geometry of NeRF can be altered without compromising the quality of reconstruction.

By computing the diagonal entries of $\Sigma$, we obtain the marginal variance vector $\boldsymbol{\sigma}=(\sigma_x,\sigma_y,\sigma_z)$. At each grid vertex, $\boldsymbol{\sigma}$ defines a spatial ellipsoid representing the region within which deformations can occur with minimal reconstruction cost. The norm of the vector $\sigma=\|\boldsymbol{\sigma}\|_2$ is a positive scalar that measures the local spatial uncertainty of the radiance field at each grid vertex.

Through this method, we can define the spatial uncertainty field $\mathcal{U}:\mathbb{R}^{3}\to\mathbb{R}^{+}$, expressed as (Fig. \ref{FIG:spatial_uncertainty}):
\begin{equation}\mathcal{U}(\mathbf{x})=\text{Trilinear}(\mathbf{x},\sigma)\end{equation}

Strictly speaking, as described above, $\mathcal{U}$ measures the uncertainty at $(1+\mathcal{D}_{\boldsymbol{\omega}})^{-1}(\mathbf{x})$, not at $\mathbf{x}$; however, for a trained SeaThru-NeRF where $\mathcal{D}_{{\boldsymbol{\omega}}^{*}}=0$, these points are effectively the same.

\section{Numerical experiment}
\subsection{Experimental setup}
\subsubsection{Dataset}
The real dataset consists of four image sets (Curasao, IUI3, Panama and JapaneseGradens) taken in three different marine environments. Prior to training, linear images were white-balanced and extreme pixel values in each channel were clipped by 0.5\% to reduce noise. The average image resolution was reduced to 900x1400, and camera poses were estimated using COLMAP \citep{COLMAP}. The synthetic data (uwSimulation) is based on the fern scene from the LLFF dataset \citep{LLFF}, with underwater simulation effects added.

\subsubsection{Implementation}
In our experiments, we set $M=256$, $\lambda=10^{-4}/M^{3}$, and iterations =1000. For the real-world and synthetic datasets, all cases except for Panama included two evaluated images, while Panama included one. We render diagonal elements of $\Sigma$ as a new volumetric data $\mathcal{U}(\mathbf{x})$, where regions with larger $\mathcal{U}(\mathbf{x})$ values indicate higher uncertainty.

Moreover, besides SeaThru-NeRF, we also conducted experiments using a smaller neural network model architecture (SeaThru-NeRF-lite) compared to SeaThru-NeRF. SeaThru-NeRF and SeaThru-NeRF-lite primarily differ in model size and training time. SeaThru-NeRF is a larger model, using approximately 23GB of VRAM, and provides optimal image quality. In contrast, SeaThru-NeRF-lite is a smaller model that requires only about 7GB of VRAM. Although the image quality is slightly lower, it still produces excellent and clear results.

\subsubsection{Metric}
We use the area under sparsification error (AUSE) \citep{ause1,ause2} to evaluate the performance of uncertainty estimation. A lower AUSE indicates that the model's uncertainty estimates are well-calibrated, meaning that higher uncertainty predictions correspond to higher actual errors. We experimented with the AUSE values related to three types of errors (MSE, MAE, RMSE).

Additionally, we also use three metrics to evaluate image quality. The structural similarity index (SSIM) \citep{SSIM} quantifies the structural similarity between two images, being sensitive to local structural changes in the image, with a range from $-1$ to $1$. Higher values indicate better image quality. Peak signal-to-noise ratio (PSNR) \citep{PSNR} evaluates image quality by calculating the MSE between the original and rendered images and converting it to a logarithmic scale. Higher values indicate better image quality. Learned perceptual image patch similarity (LPIPS) \citep{LPIPS} measures the perceptual difference between two images, prioritizing perceptual similarity. Lower LPIPS values indicate higher similarity.

\begin{figure}
	\centering
        \makebox[\textwidth][c]{
	\includegraphics[width=1.25\textwidth]{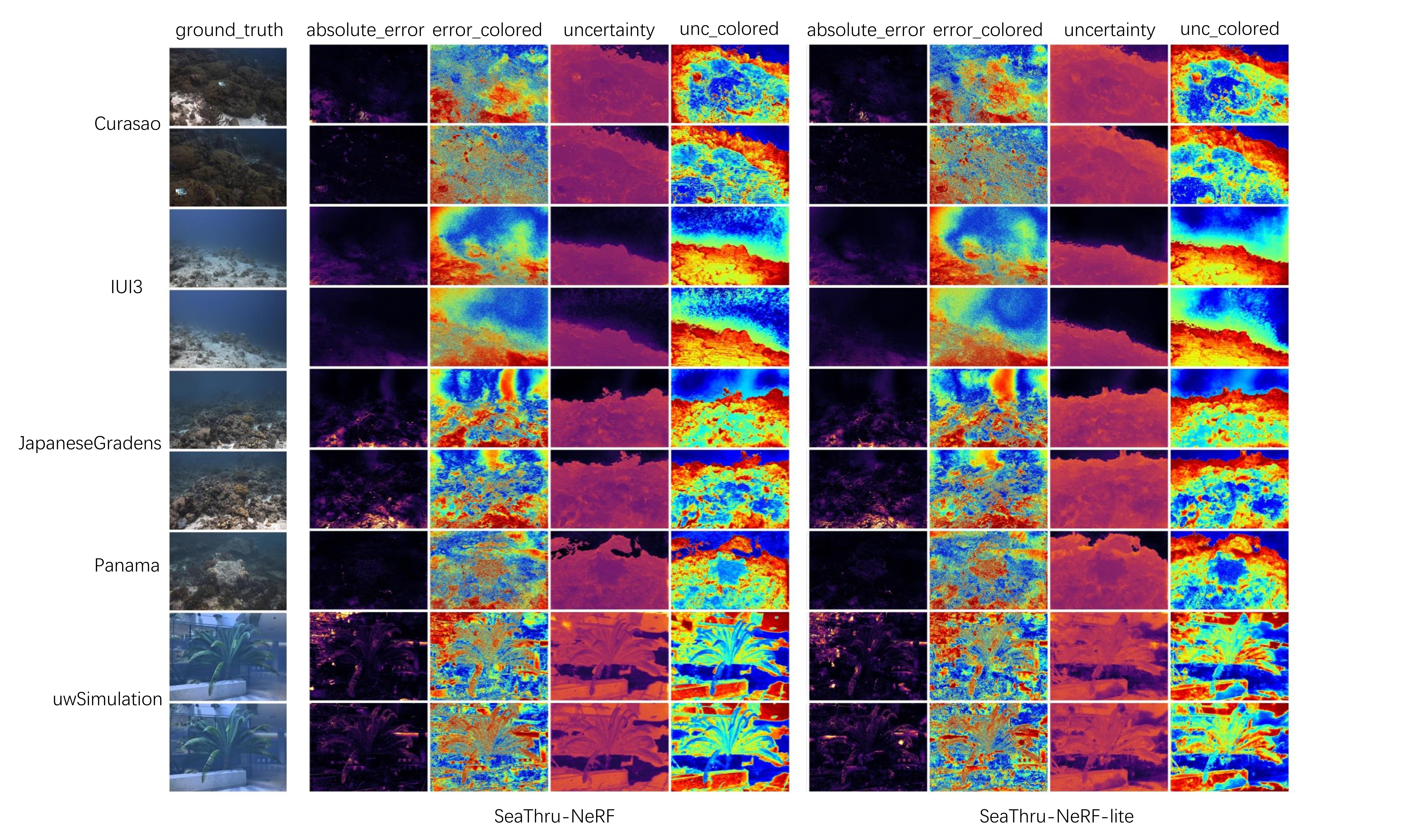}
        }
	\caption{The absolute error and uncertainty quantification of SeaThru-NeRF and SeaThru-NeRF-lite were evaluated using both real-world and synthetic datasets.}
	\label{FIG:4}
\end{figure}

\begin{table*}[ht]
    \centering
    \caption{Results of our uncertainty quantification research on SeaThru-NeRF and SeaThru-NeRF-lite.}
    \resizebox{1.2\linewidth}{!}
    {
    \begin{tabular}{@{} ccccccccccc@{} }
    \toprule
        ~ & ~ & AUSE\_MSE & AUSE\_MAE & AUSE\_RMSE  & PSNR(ours) & PSNR(base) & SSIM(ours) & SSIM(base) & LPIPS(ours) & LPIPS(base)  \\
    \midrule
        Curasao  & seathru-nerf & 0.30523 & 0.47558 & 0.3435 & 31.24606  & 31.40482  & 0.92763  & 0.93049  & 0.07061  & 0.06950   \\ 
        ~ & seathru-nerf-lite & 0.31727  & 0.51540  & 0.35525  & 31.60670  & 31.52468  & 0.93857  & 0.93716  & 0.07888  & 0.07985   \\ 
        IUI3  & seathru-nerf & 0.18834 & 0.22894 & 0.25691 & 28.76620  & 28.84461  & 0.84195  & 0.84548  & 0.16801  & 0.18165   \\ 
        ~ & seathru-nerf-lite & 0.15008  & 0.22401  & 0.22501  & 28.33715  & 28.06076  & 0.84584  & 0.83927  & 0.30336  & 0.31994   \\ 
        Panama  & seathru-nerf & 0.29347 & 0.469596 & 0.35063 & 34.05641  & 34.13037  & 0.95733  & 0.95994  & 0.04356  & 0.04148   \\ 
        ~ & seathru-nerf-lite & 0.31662  & 0.50313  & 0.37197  & 34.31783  & 34.57177  & 0.96276  & 0.95927  & 0.04249  & 0.05089   \\ 
        JapaneseGradens  & seathru-nerf & 0.04164 & 0.42559 & 0.13122 & 24.53097  & 24.61138  & 0.91968  & 0.92158  & 0.08829  & 0.08754   \\ 
        ~ & seathru-nerf-lite & 0.08290  & 0.47345  & 0.17870  & 25.11808  & 25.34451  & 0.90783  & 0.90202  & 0.11012  & 0.12594   \\ 
        uwSimulation  & seathru-nerf & 0.75512 & 0.48952 & 0.725569 & 24.92481  & 25.03314  & 0.79364  & 0.79775  & 0.20144  & 0.19730   \\ 
        ~ & seathru-nerf-lite & 0.88559  & 0.44384  & 0.76970  & 24.45243  & 25.02669  & 0.79431  & 0.80737  & 0.23303  & 0.22473   \\ 
    \bottomrule
    \end{tabular}
    }
    \label{Tab.results}
\end{table*}

\subsection{Results}
Fig. \ref{FIG:4} illustrates the uncertainty quantification results across five datasets on SeaThru-NeRF and SeaThru-NeRF-lite. The uncertainty estimates are visually conveyed using color, where the color intensity represents the level of uncertainty: bluer colors indicate higher uncertainty, and redder colors indicate lower uncertainty. In addition, as shown in Table \ref{Tab.results}, our uncertainty quantification results exhibit excellent performance across AUSE metrics. At the same time, as observed in Table \ref{Tab.results}, the PSNR, SSIM, and LPIPS metrics of the model show negligible differences compared to the original model (base). It is in line with our emphasis in Section \ref{senction4.2} on performing uncertainty quantification without significantly impacting reconstruction loss. This indicates that we can provide additional confidence information while maintaining the original reconstruction performance, which provides important technical support and guarantee for practical application. For example, in AUVs navigation, high-quality environmental reconstruction is crucial for path planning and obstacle detection. Additional uncertainty information can assist in making more cautious decisions when faced with uncertain or hazardous situations.

\subsection{Ablation study}
\subsubsection{Influence of parameter M}

\begin{figure}
	\centering
        \makebox[\textwidth][c]{
	\includegraphics[width=1.3\textwidth]{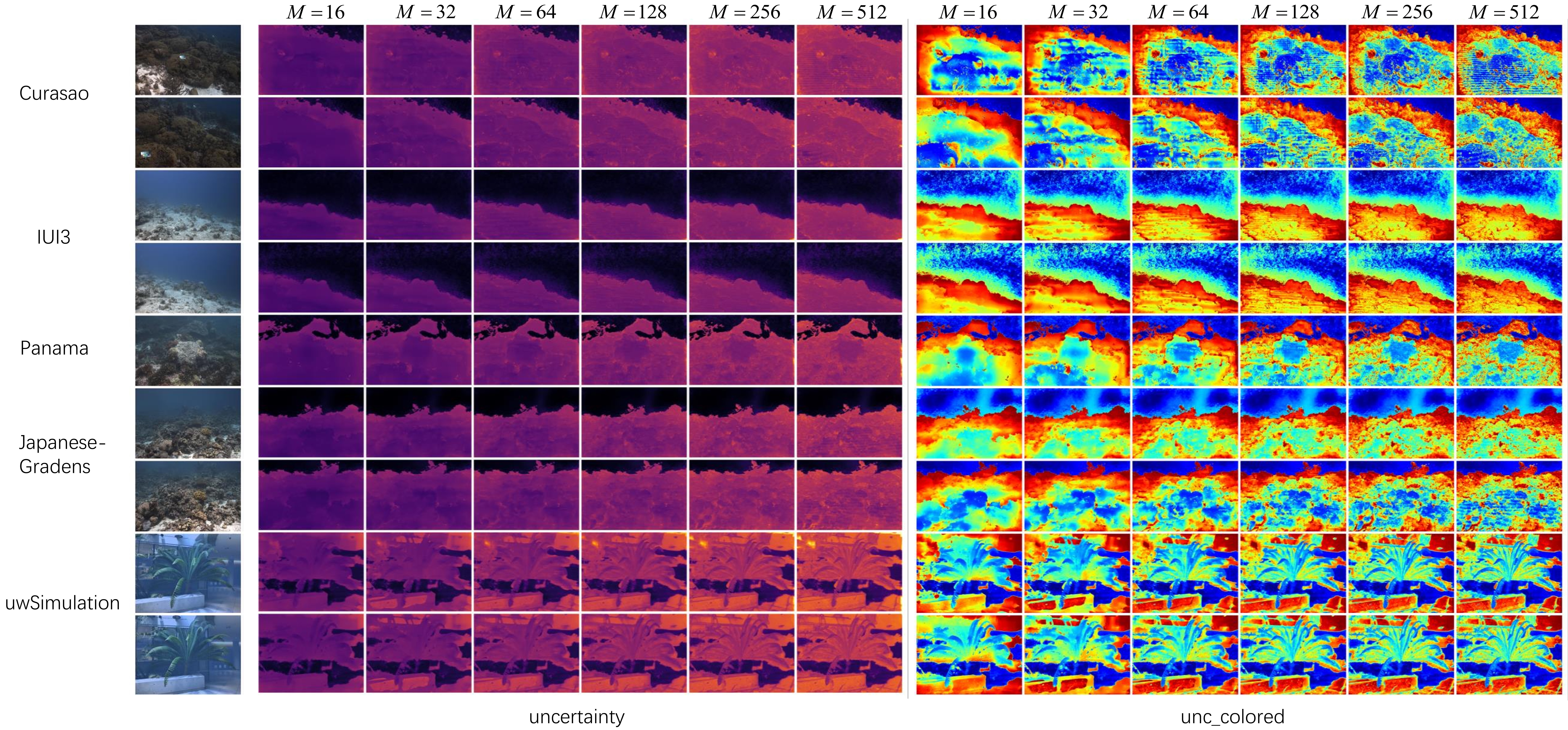}
        }
	\caption{Regarding the uncertainty under different choices of parameter $M$ on SeaThru-NeRF.}
	\label{FIG:seathru-nerf-M}
\end{figure}

\begin{figure}
	\centering
        \makebox[\textwidth][c]{
	\includegraphics[width=1.3\textwidth]{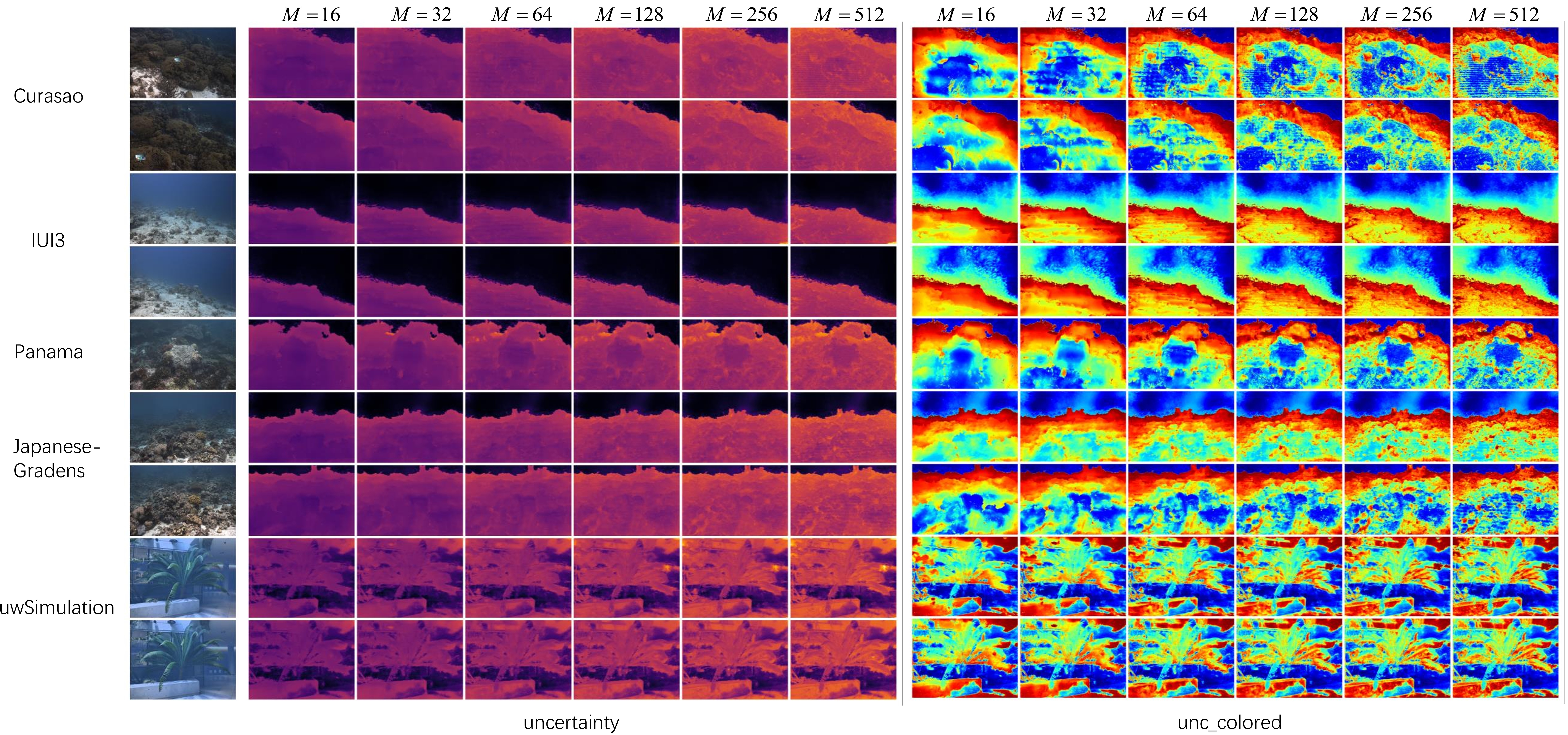}
        }
	\caption{Regarding the uncertainty under different choices of parameter $M$ on SeaThru-NeRF-lite.}
	\label{FIG:seathru-nerf-lite-M}
\end{figure}

In Fig. \ref{FIG:seathru-nerf-M}, Fig. \ref{FIG:seathru-nerf-lite-M} and Table \ref{Tab.1}, we present the uncertainty estimates at different parameter $M$ (i.e., grid size). $M$ determines the granularity of the spatial segmentation, thereby influencing the model's ability to capture scene details and the accuracy of uncertainty estimation. Analyzing uncertainty estimates at different grid sizes allows for a deeper understanding of the influence of parameter $M$.

\begin{table*}[ht]
    \centering
    \caption{Ablation study of parameter $M$.}
    \resizebox{1.2\linewidth}{!}
    {
    \begin{tabular}{@{\extracolsep\fill}llccccccccc}
    \toprule
    & & \multicolumn{3}{@{}c@{}}{$M=16$} & \multicolumn{3}{@{}c@{}}{$M=32$} & \multicolumn{3}{@{}c@{}}{$M=64$} \\\cmidrule(lr){3-5}\cmidrule(lr){6-8}\cmidrule(lr){9-11}%
        ~ & ~ & AUSE\_MSE & AUSE\_MAE & AUSE\_RMSE & AUSE\_MSE & AUSE\_MAE & AUSE\_RMSE & AUSE\_MSE & AUSE\_MAE & AUSE\_RMSE  \\ 
    \midrule
        Curasao & seathru-nerf & 0.31296 & 0.46118 & 0.34732 & 0.30963 & 0.46208 & 0.34561 & 0.31188 & 0.47111 & 0.34674  \\ 
        ~ & seathru-nerf-lite & 0.31954 & 0.47982 & 0.35705 & 0.32095 & 0.4935 & 0.3573 & 0.32056 & 0.50304 & 0.35708  \\ 
        IUI3 & seathru-nerf & 0.188496 & 0.23426 & 0.25706 & 0.18846 & 0.23242 & 0.25702 & 0.18822 & 0.23021 & 0.25684  \\ 
        ~ & seathru-nerf-lite & 0.15069 & 0.22814 & 0.22551 & 0.15035 & 0.22653 & 0.22525 & 0.15008 & 0.22488 & 0.22504  \\ 
        Panama & seathru-nerf & 0.29554 & 0.48057 & 0.35222 & 0.29417 & 0.47178 & 0.35118 & 0.29385 & 0.47117 & 0.35092  \\ 
        ~ & seathru-nerf-lite & 0.31977 & 0.53474 & 0.37435 & 0.3186 & 0.51457 & 0.37348 & 0.31747 & 0.51243 & 0.37262  \\ 
        JapaneseGradens & seathru-nerf & 0.04112 & 0.41962 & 0.13046 & 0.04146 & 0.40946 & 0.13097 & 0.04148 & 0.40756 & 0.13099  \\ 
        ~ & seathru-nerf-lite & 0.08387 & 0.46099 & 0.17996 & 0.08352 & 0.4519 & 0.17947 & 0.08331 & 0.4529 & 0.17921  \\ 
        uwSimulation & seathru-nerf & 0.74244 & 0.46726 & 0.71863 & 0.75712 & 0.46848 & 0.72628 & 0.75442 & 0.48468 & 0.72511  \\ 
        ~ & seathru-nerf-lite & 0.85303 & 0.43378 & 0.75572 & 0.86718 & 0.43215 & 0.76192 & 0.88133 & 0.43807 & 0.76797  \\ 
    \bottomrule
    \\
    \toprule
    & & \multicolumn{3}{@{}c@{}}{$M=128$} & \multicolumn{3}{@{}c@{}}{$M=256$} & \multicolumn{3}{@{}c@{}}{$M=512$} \\\cmidrule(lr){3-5}\cmidrule(lr){6-8}\cmidrule(lr){9-11}%
        ~ & ~ & AUSE\_MSE & AUSE\_MAE & AUSE\_RMSE & AUSE\_MSE & AUSE\_MAE & AUSE\_RMSE & AUSE\_MSE & AUSE\_MAE & AUSE\_RMSE  \\ 
    \midrule
        Curasao & seathru-nerf & 0.30743 & 0.47222 & 0.3446 & 0.30523 & 0.47558 & 0.3435 & 0.30819 & 0.50488 & 0.34504  \\ 
        ~ & seathru-nerf-lite & 0.31838 & 0.51017 & 0.35592 & 0.31727 & 0.515399 & 0.35525 & 0.31773 & 0.514298 & 0.35596  \\ 
        IUI3 & seathru-nerf & 0.18826 & 0.22971 & 0.25686 & 0.18834 & 0.22894 & 0.25691 & 0.18834 & 0.23043 & 0.25693  \\ 
        ~ & seathru-nerf-lite & 0.15006 & 0.22467 & 0.22501 & 0.15008 & 0.22401 & 0.22501 & 0.14999 & 0.22435 & 0.22496  \\ 
        Panama & seathru-nerf & 0.29369 & 0.47242 & 0.35079 & 0.29347 & 0.469596 & 0.35063 & 0.29354 & 0.46585 & 0.35068  \\ 
        ~ & seathru-nerf-lite & 0.31695 & 0.50764 & 0.37222 & 0.31662 & 0.50313 & 0.37197 & 0.31623 & 0.49484 & 0.37166  \\ 
        JapaneseGradens & seathru-nerf & 0.04182 & 0.42031 & 0.13151 & 0.04164 & 0.42559 & 0.13122 & 0.042 & 0.4463 & 0.13178  \\ 
        ~ & seathru-nerf-lite & 0.08336 & 0.4627 & 0.17928 & 0.0829 & 0.47345 & 0.1787 & 0.08328 & 0.48051 & 0.17919  \\ 
        uwSimulation & seathru-nerf & 0.75756 & 0.48492 & 0.72675 & 0.75512 & 0.48952 & 0.725569 & 0.75617 & 0.50209 & 0.72618  \\ 
        ~ & seathru-nerf-lite & 0.88521 & 0.44128 & 0.76963 & 0.88559 & 0.44384 & 0.7697 & 0.88561 & 0.45851 & 0.76972  \\ 
    \bottomrule
    \end{tabular}
    }
    \label{Tab.1}
\end{table*}

Through the Table \ref{Tab.1}, it can be observed that using extremely low grid sizes (e.g., $M=16$) leads to insufficient uncertainty estimates. Low grid sizes mean that spatial segmentation is coarser, resulting in a loss of detailed information in the image. This loss of information prevents the model from adequately capturing the complex features of the scene, especially in regions where high-frequency variations and complex geometries are present. In this case, the model may exhibit overconfidence, thereby underestimating the actual uncertainty. Therefore, although the computational cost is lower at low grid sizes, the uncertainty estimates are also less accurate and cannot provide reliable confidence information.

As the grid size increases, the effectiveness of uncertainty estimation gradually improves. When the grid size reaches a certain level (e.g., $M=256$), the uncertainty estimation results achieve optimal performance. A higher grid size implies finer spatial segmentation, which helps capture more image details and scene features. In this case, the model can more accurately identify and quantify uncertainties, particularly in complex and variable regions. Therefore, uncertainty estimation at higher grid sizes is more reliable and better reflects the uncertainties in the model's predictions. 

Specifically, it is experimentally found that the uncertainty estimation can reach the optimal equilibrium point at a moderate grid size (e.g., $M=256$). With this grid size, the model is able to make full use of the spatial segmentation accuracy to accurately capture image details and scene features, thus providing high-quality uncertainty estimation. At the same time, the computational resources and time costs are within acceptable range, achieving an optimal balance between performance and efficiency. In other words, a moderate $M$ value (e.g., $M=256$) can provide high-quality uncertainty estimation while ensuring computational efficiency.

However, when the grid size continues to increase (e.g., $M=512$), the benefits of uncertainty estimation start to diminish. This phenomenon can be attributed to several factors. First, a significantly higher grid size brings substantial increases in computational resources and time. Although the increased spatial segmentation precision can capture more detailed information, these additional details do not significantly improve the accuracy of uncertainty estimation. In other words, beyond a certain grid size, the increased computational complexity and time cost do not yield corresponding performance gains, potentially leading to resource wastage. Additionally, excessively high grid sizes may introduce extra noise and uncertainty. Despite the finer spatial segmentation, these extra partitions may introduce more noise in complex scenes, thereby affecting the model's uncertainty estimation.

In summary, the parameter $M$ plays a crucial role in our research. An appropriate grid size not only improves the accuracy of uncertainty estimation, but also finds an optimal balance between computational resources and time cost. This observation provides an important reference for parameter selection in practical applications.

\subsubsection{Influence of parameter lambda}

In uncertainty quantification tasks, the choice of the regularization parameter $\lambda$ is often a critical factor. To validate the sensitivity and robustness of the $\lambda$ parameter selection, we conducted ablation experiments, selecting multiple different $\lambda$ values and observing the changes in the AUSE metric.

\begin{figure}
	\centering
        \makebox[\textwidth][c]{
	\includegraphics[width=1.3\textwidth]{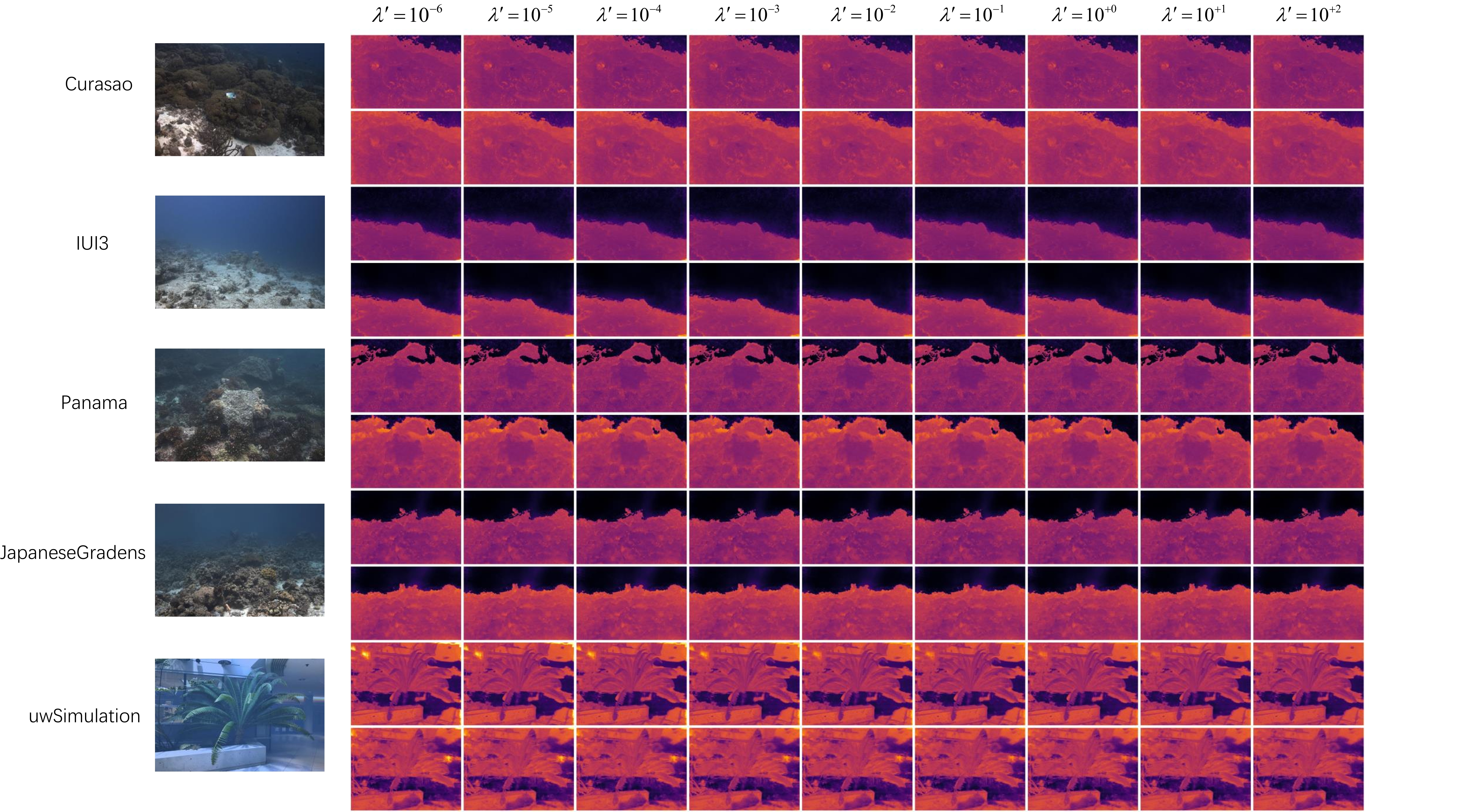}
        }
	\caption{Uncertainty of ablation study on choice of $\lambda$ where $\lambda={\lambda^{\prime}}/256^{3}$. And the odd-numbered rows correspond to SeaThru-NeRF and even-numbered rows correspond to SeaThru-NeRF-lite.}
	\label{lambda1}
\end{figure}
\begin{figure}
	\centering
        \makebox[\textwidth][c]{
	\includegraphics[width=1.3\textwidth]{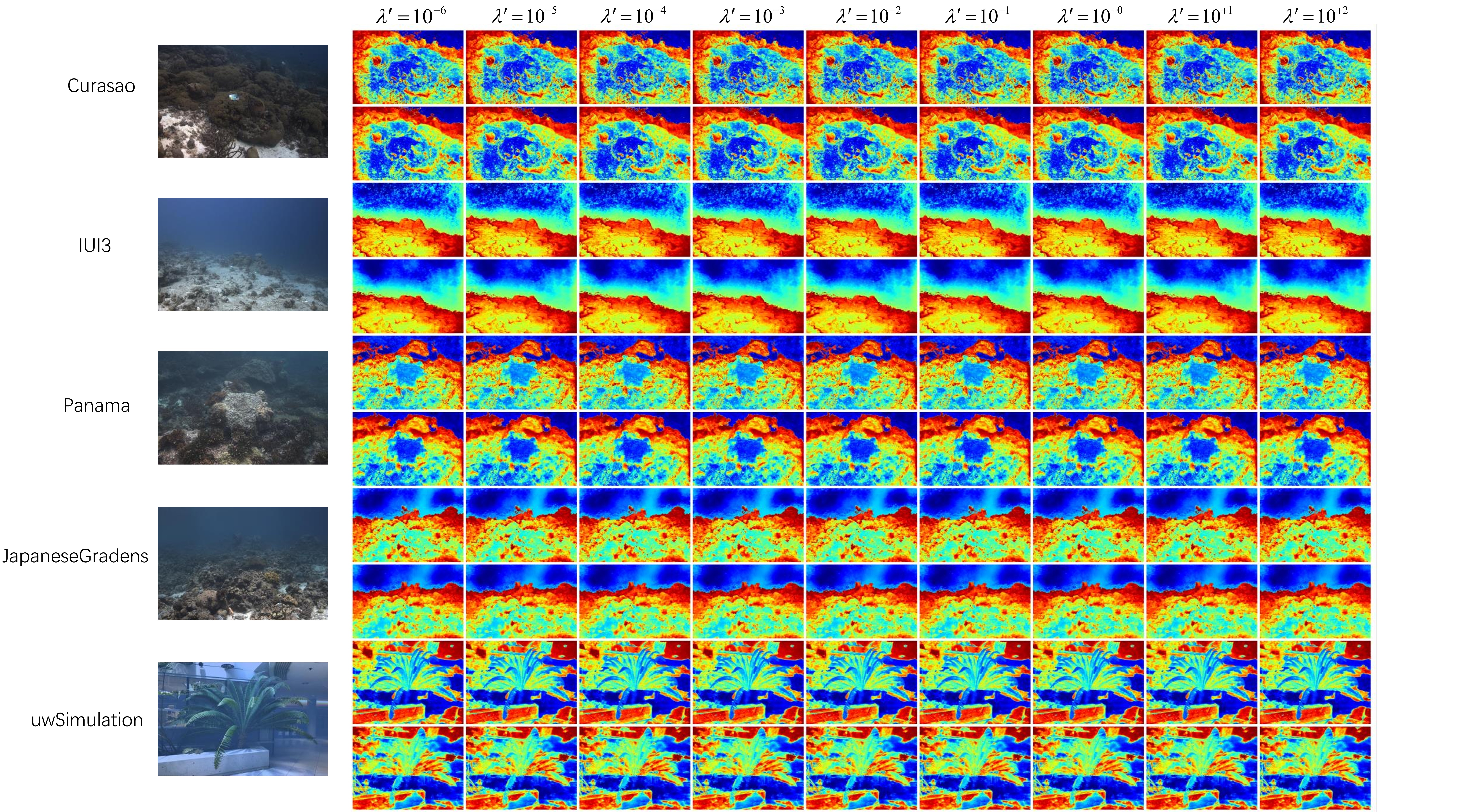}
        }
	\caption{Colored uncertainty of ablation study on choice of $\lambda$ where $\lambda={\lambda^{\prime}}/256^{3}$. And the odd-numbered rows correspond to SeaThru-NeRF and even-numbered rows correspond to SeaThru-NeRF-lite.}
	\label{lambda2}
\end{figure}

\begin{table*}[ht]
    \centering
    \caption{Ablation study of parameter $\lambda$.}
    \resizebox{1.2\linewidth}{!}{
    \begin{tabular}{@{} ccccccccccc@{} }
    \toprule
    & & \multicolumn{3}{@{}c@{}}{$\lambda=10^{-6}/256^{3}$} & \multicolumn{3}{@{}c@{}}{$\lambda=10^{-5}/256^{3}$} & \multicolumn{3}{@{}c@{}}{$\lambda=10^{-4}/256^{3}$} \\\cmidrule(lr){3-5}\cmidrule(lr){6-8}\cmidrule(lr){9-11}%
        ~ & ~ & AUSE\_MSE & AUSE\_MAE & AUSE\_RMSE & AUSE\_MSE & AUSE\_MAE & AUSE\_RMSE & AUSE\_MSE & AUSE\_MAE & AUSE\_RMSE  \\
    \midrule
        Curasao & seathru-nerf & 0.305097  & 0.475915  & 0.343446  & 0.305877  & 0.475641  & 0.343836  & 0.305227  & 0.475575  & 0.343502   \\ 
        ~ & seathru-nerf-lite & 0.317300  & 0.514976  & 0.355272  & 0.317226  & 0.515023  & 0.355213  & 0.317270  & 0.515399  & 0.355254   \\ 
        IUI3 & seathru-nerf & 0.188341  & 0.228918  & 0.256907  & 0.188336  & 0.228961  & 0.256903  & 0.188344  & 0.228937  & 0.256910   \\ 
        ~ & seathru-nerf-lite & 0.150081  & 0.224004  & 0.225014  & 0.150080  & 0.223975  & 0.225012  & 0.150083  & 0.224009  & 0.225015   \\ 
        Panama & seathru-nerf & 0.293607  & 0.469783  & 0.350732  & 0.293460  & 0.469740  & 0.350620  & 0.293474  & 0.469596  & 0.350631   \\ 
        ~ & seathru-nerf-lite & 0.316573  & 0.503292  & 0.371936  & 0.316534  & 0.503354  & 0.371907  & 0.316621  & 0.503130  & 0.371973   \\ 
        JapaneseGradens & seathru-nerf & 0.041607  & 0.422099  & 0.131177  & 0.041624  & 0.423969  & 0.131204  & 0.041635  & 0.425591  & 0.131222   \\ 
        ~ & seathru-nerf-lite & 0.082522  & 0.470332  & 0.178238  & 0.082702  & 0.472465  & 0.178463  & 0.082895  & 0.473451  & 0.178703   \\ 
        uwSimulation & seathru-nerf & 0.758189  & 0.490118  & 0.727131  & 0.756315  & 0.489870  & 0.726176  & 0.755125  & 0.489520  & 0.725569   \\ 
        ~ & seathru-nerf-lite & 0.885232  & 0.443938  & 0.769525  & 0.885482  & 0.443795  & 0.769653  & 0.885589  & 0.443840  & 0.769704   \\ 
    \bottomrule
        \\ 
    \toprule
    & & \multicolumn{3}{@{}c@{}}{$\lambda=10^{-3}/256^{3}$} & \multicolumn{3}{@{}c@{}}{$\lambda=10^{-2}/256^{3}$} & \multicolumn{3}{@{}c@{}}{$\lambda=10^{-1}/256^{3}$} \\\cmidrule(lr){3-5}\cmidrule(lr){6-8}\cmidrule(lr){9-11}
        ~ & ~ & AUSE\_MSE & AUSE\_MAE & AUSE\_RMSE & AUSE\_MSE & AUSE\_MAE & AUSE\_RMSE & AUSE\_MSE & AUSE\_MAE & AUSE\_RMSE  \\
    \midrule
        Curasao & seathru-nerf & 0.305245  & 0.475608  & 0.343517  & 0.304875  & 0.475516  & 0.343322  & 0.305049  & 0.475232  & 0.343407   \\ 
        ~ & seathru-nerf-lite & 0.317319  & 0.514914  & 0.355267  & 0.317206  & 0.515218  & 0.355213  & 0.317342  & 0.514882  & 0.355284   \\ 
        IUI3 & seathru-nerf & 0.188341  & 0.228972  & 0.256908  & 0.188351  & 0.228999  & 0.256916  & 0.188357  & 0.229069  & 0.256920   \\ 
        ~ & seathru-nerf-lite & 0.150083  & 0.223981  & 0.225015  & 0.150082  & 0.224050  & 0.225014  & 0.150082  & 0.224060  & 0.225014   \\ 
        Panama & seathru-nerf & 0.293440  & 0.469665  & 0.350604  & 0.293421  & 0.469665  & 0.350590  & 0.293327  & 0.469757  & 0.350516   \\ 
        ~ & seathru-nerf-lite & 0.316556  & 0.503609  & 0.371923  & 0.316557  & 0.503048  & 0.371922  & 0.316608  & 0.502941  & 0.371962   \\ 
        JapaneseGradens & seathru-nerf & 0.041650  & 0.427574  & 0.131244  & 0.041664  & 0.429283  & 0.131266  & 0.041687  & 0.431087  & 0.131301   \\ 
        ~ & seathru-nerf-lite & 0.083091  & 0.473875  & 0.178947  & 0.083290  & 0.474727  & 0.179195  & 0.083499  & 0.475687  & 0.179453   \\ 
        uwSimulation & seathru-nerf & 0.753546  & 0.489448  & 0.724759  & 0.752280  & 0.489259  & 0.724111  & 0.751624  & 0.489032  & 0.723774   \\ 
        ~ & seathru-nerf-lite & 0.885756  & 0.443877  & 0.769787  & 0.885542  & 0.443982  & 0.769699  & 0.884586  & 0.443875  & 0.769307   \\ 
    \bottomrule
        \\ 
    \toprule
    & & \multicolumn{3}{@{}c@{}}{$\lambda=10^{+0}/256^{3}$} & \multicolumn{3}{@{}c@{}}{$\lambda=10^{+1}/256^{3}$} & \multicolumn{3}{@{}c@{}}{$\lambda=10^{+2}/256^{3}$} \\\cmidrule(lr){3-5}\cmidrule(lr){6-8}\cmidrule(lr){9-11}
        ~ & ~ & AUSE\_MSE & AUSE\_MAE & AUSE\_RMSE & AUSE\_MSE & AUSE\_MAE & AUSE\_RMSE & AUSE\_MSE & AUSE\_MAE & AUSE\_RMSE  \\
    \midrule
        Curasao & seathru-nerf & 0.306121  & 0.475468  & 0.343931  & 0.305626  & 0.475344  & 0.343692  & 0.306133  & 0.475322  & 0.343947   \\ 
        ~ & seathru-nerf-lite & 0.317286  & 0.515510  & 0.355253  & 0.317258  & 0.515088  & 0.355234  & 0.317231  & 0.515209  & 0.355195   \\ 
        IUI3 & seathru-nerf & 0.188358  & 0.229056  & 0.256921  & 0.188373  & 0.229157  & 0.256932  & 0.188397  & 0.229244  & 0.256949   \\ 
        ~ & seathru-nerf-lite & 0.150068  & 0.224163  & 0.225000  & 0.150075  & 0.224310  & 0.224998  & 0.150094  & 0.224451  & 0.224999   \\ 
        Panama & seathru-nerf & 0.293328  & 0.469565  & 0.350517  & 0.293288  & 0.469674  & 0.350486  & 0.293303  & 0.469529  & 0.350498   \\ 
        ~ & seathru-nerf-lite & 0.316509  & 0.503183  & 0.371886  & 0.316584  & 0.502911  & 0.371944  & 0.316537  & 0.503446  & 0.371908   \\ 
        JapaneseGradens & seathru-nerf & 0.041713  & 0.433563  & 0.131342  & 0.041735  & 0.436043  & 0.131374  & 0.041759  & 0.437183  & 0.131410   \\ 
        ~ & seathru-nerf-lite & 0.083736  & 0.476267  & 0.179745  & 0.083949  & 0.476716  & 0.180006  & 0.084140  & 0.476693  & 0.180240   \\ 
        uwSimulation & seathru-nerf & 0.750872  & 0.488936  & 0.723385  & 0.749251  & 0.488403  & 0.722554  & 0.746942  & 0.488111  & 0.721370   \\ 
        ~ & seathru-nerf-lite & 0.884037  & 0.444256  & 0.769081  & 0.877413  & 0.444275  & 0.766256  & 0.873298  & 0.443020  & 0.764476   \\
    \bottomrule
    \end{tabular}
    }
    \label{Tab.lambda}
\end{table*} 

From Fig. \ref{lambda1} and Fig. \ref{lambda2}, we can observe that the change in uncertainty is not significant. And at the same time, the experimental results in Table \ref{Tab.lambda} show that, despite $\lambda$ values spanning several orders of magnitude, the range of changes in the AUSE metric is very limited. This indicates that our method maintains good performance over a wide range of $\lambda$ values, and the minimal changes in the AUSE metric further demonstrate the robustness and stability of the method. The method shows stable performance when handling different levels of regularization, ensuring that the model can still provide reliable results under different settings. 

The insensitivity to the choice of the $\lambda$ value provides significant convenience for practical applications. Since the model maintains good performance across a relatively wide range of $\lambda$ values, users do not need to fine-tune the $\lambda$ value excessively during actual operation. This characteristic greatly simplifies the use of the model and reduces the cost of parameter optimization in different application scenarios. Meanwhile, the robustness of the model is enhanced, enabling it to operate reliably in various complex and dynamic environments. This feature is particularly important for practical application scenarios that require rapid deployment and real-time response, as it reduces the difficulty of adapting the model to different datasets and task requirements, enhancing the method's versatility and operability. Consequently, it provides a solid technical foundation for broad application.

\subsubsection{Influence of iterations}
In experiments, the number of iterations determines the duration of the optimization process. More iterations typically mean more opportunities for the model to update parameters, which may lead to finding better solutions. Therefore, higher iteration counts may improve model accuracy and stability. However, excessively high iteration counts can also significantly increase training time, affecting computational efficiency. To analyze the impact of iteration counts in depth, we conducted experiments with different iteration counts.

\begin{figure}
	\centering
        \makebox[\textwidth][c]{
	\includegraphics[width=1.25\textwidth]{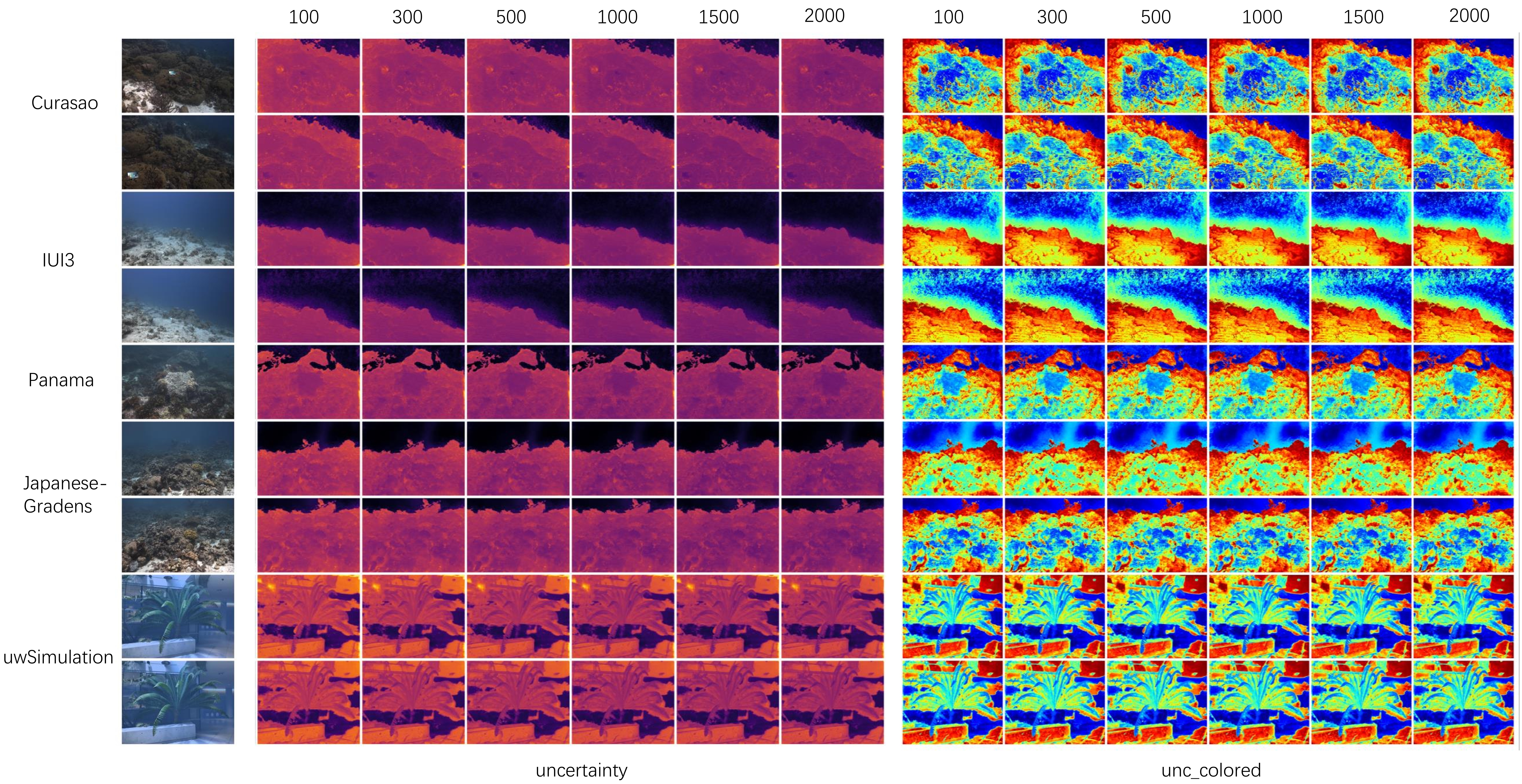}
        }
	\caption{Ablation study of parameter iterations on SeaThru-NeRF.}
        \label{iterations1}
\end{figure}
\begin{figure}
	\centering
        \makebox[\textwidth][c]{
	\includegraphics[width=1.25\textwidth]{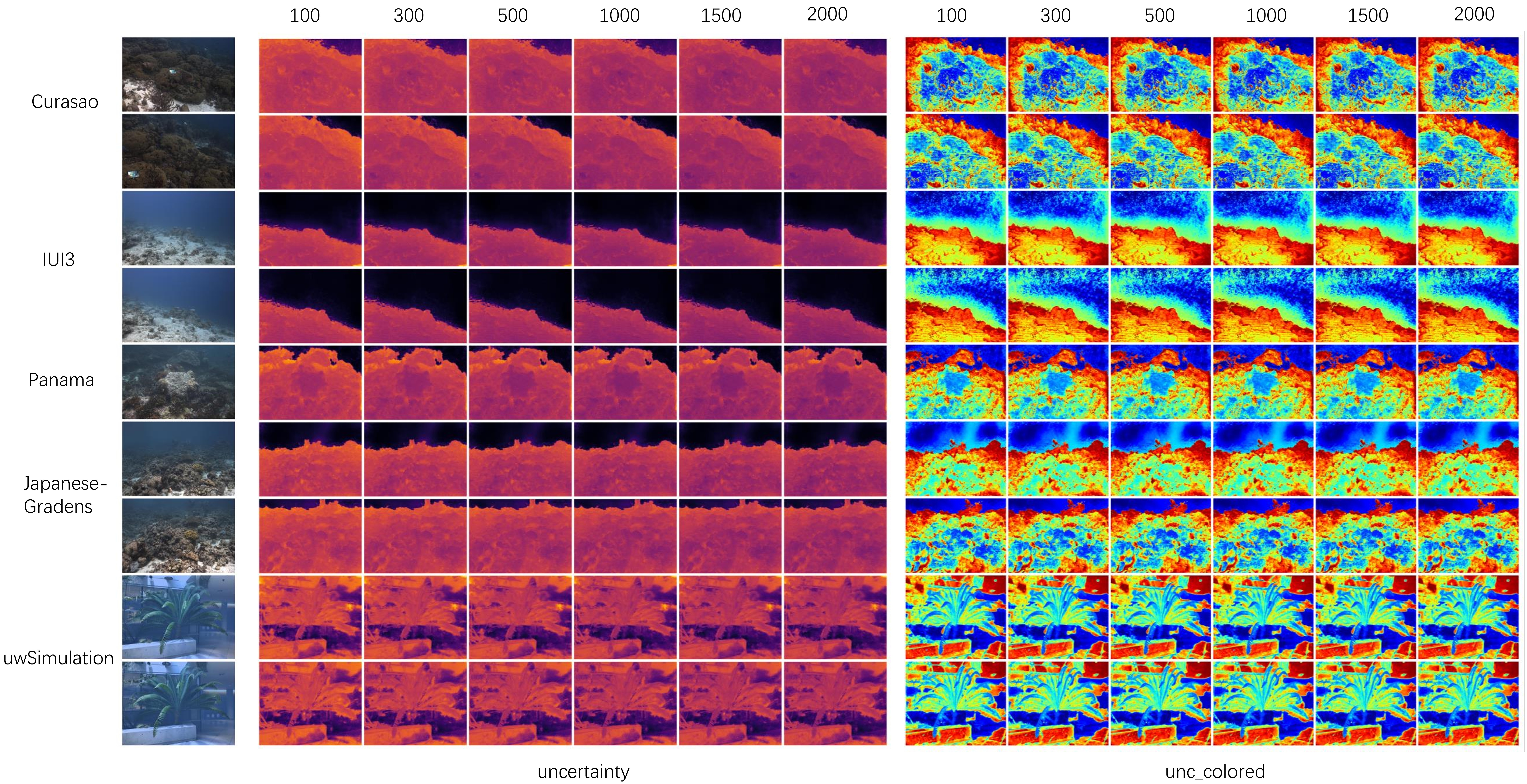}
        }
	\caption{Ablation study of parameter iterations on SeaThru-NeRF-lite.}
        \label{iterations2}
\end{figure}
\begin{table*}[ht]
    \centering
    \caption{Ablation study of parameter iterations.}\label{iterations}
    \resizebox{1.2\textwidth}{!}{
    \begin{tabular}{@{} ccccccccccc@{} }
    \toprule
    & & \multicolumn{3}{@{}c@{}}{iterations=100} & \multicolumn{3}{@{}c@{}}{iterations=200} & \multicolumn{3}{@{}c@{}}{iterations=300} \\\cmidrule(lr){3-5}\cmidrule(lr){6-8}\cmidrule(lr){9-11}
        ~ & ~ & AUSE\_MSE & AUSE\_MAE & AUSE\_RMSE & AUSE\_MSE & AUSE\_MAE & AUSE\_RMSE & AUSE\_MSE & AUSE\_MAE & AUSE\_RMSE  \\
    \midrule
        Curasao & seathru-nerf & 0.305659  & 0.473639  & 0.343713  & 0.305385  & 0.475084  & 0.343568  & 0.304905  & 0.475541  & 0.343346   \\ 
        ~ & seathru-nerf-lite & 0.317328  & 0.514631  & 0.355282  & 0.317231  & 0.514994  & 0.355215  & 0.317237  & 0.514179  & 0.355210   \\ 
        IUI3 & seathru-nerf & 0.188374  & 0.228354  & 0.256929  & 0.188378  & 0.228885  & 0.256934  & 0.188370  & 0.228906  & 0.256928   \\ 
        ~ & seathru-nerf-lite & 0.150142  & 0.223699  & 0.225054  & 0.150125  & 0.223928  & 0.225044  & 0.150100  & 0.223972  & 0.225026   \\ 
        Panama & seathru-nerf & 0.293459  & 0.470040  & 0.350619  & 0.293425  & 0.469367  & 0.350593  & 0.293382  & 0.469639  & 0.350559   \\ 
        ~ & seathru-nerf-lite & 0.316639  & 0.502895  & 0.371987  & 0.316584  & 0.503721  & 0.371944  & 0.316566  & 0.502956  & 0.371930   \\ 
        JapaneseGradens & seathru-nerf & 0.041648  & 0.425957  & 0.131242  & 0.041638  & 0.425930  & 0.131226  & 0.041625  & 0.426061  & 0.131207   \\ 
        ~ & seathru-nerf-lite & 0.083123  & 0.473766  & 0.178990  & 0.083001  & 0.473642  & 0.178834  & 0.082956  & 0.473534  & 0.178781   \\ 
        uwSimulation & seathru-nerf & 0.752581  & 0.486382  & 0.724259  & 0.752934  & 0.488215  & 0.724438  & 0.755410  & 0.488788  & 0.725712   \\ 
        ~ & seathru-nerf-lite & 0.885621  & 0.436375  & 0.769709  & 0.885757  & 0.440941  & 0.769786  & 0.885783  & 0.442063  & 0.769803   \\ 
    \bottomrule
        \\ 
    \toprule
    & & \multicolumn{3}{@{}c@{}}{iterations=1000} & \multicolumn{3}{@{}c@{}}{iterations=1500} & \multicolumn{3}{@{}c@{}}{iterations=2000} \\\cmidrule(lr){3-5}\cmidrule(lr){6-8}\cmidrule(lr){9-11}
        ~ & ~ & AUSE\_MSE & AUSE\_MAE & AUSE\_RMSE & AUSE\_MSE & AUSE\_MAE & AUSE\_RMSE & AUSE\_MSE & AUSE\_MAE & AUSE\_RMSE  \\
    \midrule
        Curasao & seathru-nerf & 0.305227  & 0.475575  & 0.343502  & 0.305797  & 0.475731  & 0.343802  & 0.304581  & 0.475949  & 0.343189   \\ 
        ~ & seathru-nerf-lite & 0.317270  & 0.515399  & 0.355254  & 0.317271  & 0.515034  & 0.355237  & 0.317286  & 0.515187  & 0.355257   \\ 
        IUI3 & seathru-nerf & 0.188344  & 0.228937  & 0.256910  & 0.188335  & 0.229022  & 0.256904  & 0.188324  & 0.229073  & 0.256896   \\ 
        ~ & seathru-nerf-lite & 0.150083  & 0.224009  & 0.225015  & 0.150066  & 0.224009  & 0.225003  & 0.150056  & 0.224019  & 0.224996   \\ 
        Panama & seathru-nerf & 0.293474  & 0.469596  & 0.350631  & 0.293464  & 0.469693  & 0.350622  & 0.293543  & 0.469767  & 0.350681   \\ 
        ~ & seathru-nerf-lite & 0.316621  & 0.503130  & 0.371973  & 0.316590  & 0.503392  & 0.371949  & 0.316549  & 0.503300  & 0.371918   \\ 
        JapaneseGradens & seathru-nerf & 0.041635  & 0.425591  & 0.131222  & 0.041635  & 0.425580  & 0.131222  & 0.041636  & 0.425358  & 0.131223   \\ 
        ~ & seathru-nerf-lite & 0.082895  & 0.473451  & 0.178703  & 0.082863  & 0.473325  & 0.178665  & 0.082821  & 0.473129  & 0.178611   \\ 
        uwSimulation & seathru-nerf & 0.755125  & 0.489520  & 0.725569  & 0.755997  & 0.490056  & 0.726017  & 0.756580  & 0.490418  & 0.726316   \\ 
        ~ & seathru-nerf-lite & 0.885589  & 0.443840  & 0.769704  & 0.885517  & 0.444950  & 0.769669  & 0.885512  & 0.445619  & 0.769669   \\
    \bottomrule
    \end{tabular}}
\end{table*}

From the experimental results in Fig. \ref{iterations1}, Fig. \ref{iterations2} and Table \ref{iterations}, we observed certain regularities in the influence of iteration counts on uncertainty quantification metrics for both real and synthetic datasets. Generally speaking, increasing the number of iterations improves performance to some extent, but fluctuations in performance were also observed under certain conditions.

In real datasets, the model's performance across various metrics gradually improves with increasing iteration counts until optimal performance is achieved, indicating effective capture of dataset complexity and thorough training. However, in synthetic datasets, the  model often performs better in initial iterations than later ones, suggesting that early iterations adequately capture the dataset's features, while excessive iterations may lead to overfitting.

In the synthetic dataset, the SeaThru-NeRF model has the smallest AUSE\_MSE, AUSE\_MAE, and AUSE\_RMSE metrics at an iteration number of 100, indicating that the model is able to converge quickly and achieve the best performance at a smaller number of iterations. Also, the SeaThru-NeRF-lite model has the smallest AUSE\_MAE metrics at an iteration number of 100, indicating that the model is able to perform well in some metrics during the initial iteration phase. However, as the number of iterations increases, the performance metrics do not continue to improve and even deteriorate. This phenomenon may be caused by several factors. 

Firstly, the rapid initial convergence of the model may be due to the fact that within a small number of iterations, the model is able to efficiently tune the parameters to quickly find a better locally optimal solution, which results in excellent performance metrics. For synthetic datasets, the initial fast tuning may be sufficient to capture the main features of the data and achieve better performance. However, as the number of iterations increases, the model may begin to overfit the training data, resulting in performance metrics that no longer improve or even deteriorate. This is particularly evident in the SeaThru-NeRF model, suggesting that after a certain threshold number of iterations, the model overfits the details of the training data and instead reduces its ability to generalise over the test data.

Moreover, synthetic datasets often possess specific patterns and characteristics captured by the model in early iterations. For the SeaThru-NeRF-lite, minimal AUSE\_MSE and AUSE\_RMSE were observed at 2000 iterations, indicating improved adaptation to dataset characteristics over extended training, despite its overall performance being inferior to SeaThru-NeRF. This reflects the impact of model scale differences during training, where the larger SeaThru-NeRF model benefits from rapid convergence due to its complex structure, while the smaller SeaThru-NeRF-lite model may require more iterations to achieve comparable results.

Overall, the impact of iteration counts on model performance exhibits certain regularities, with longer iterations generally enhancing performance and robustness. However, optimal iteration counts vary depending on dataset and model architecture, underscoring the importance of selecting suitable iteration counts based on specific application scenarios and data characteristics. Rational iteration count selection can enhance model performance and stability, thereby providing more reliable results in practical applications.

\subsection{Discussion}
\subsubsection{Influence of model architecture}

From Figs. \ref{FIG:4}-\ref{iterations2} and Tables \ref{Tab.results}-\ref{iterations}, it can be noted that SeaThru-NeRF-lite is slightly inferior to SeaThru-NeRF in terms of uncertainty estimate. This discrepancy is due to the combination of multiple factors.

Firstly, SeaThru-NeRF is a larger model, providing higher capacity and representational power. Consequently, it captures more details and complex features, leading to higher quality images and more accurate uncertainty estimates. In contrast, SeaThru-NeRF-lite, being smaller in size with limited parameters, has weaker representational power and detail capture, resulting in slightly inferior uncertainty estimates. 

Secondly, SeaThru-NeRF benefits from longer training times and deeper neural network layers during training, better fitting the data distribution and reducing model uncertainty. SeaThru-NeRF-lite is simplified in terms of iterations and network architecture, leading to its comparatively lower performance in uncertainty quantification.

Thirdly, the larger model capacity allows SeaThru-NeRF to better fit the training data, demonstrating more stable and superior performance across different parameter settings. Specifically, SeaThru-NeRF effectively utilizes its higher parameter capacity to optimize the model when facing various $M$, $\lambda$, and iterations settings, excelling in AUSE\_MSE, AUSE\_MAE, and AUSE\_RMSE metrics. In contrast, SeaThru-NeRF-lite, with its smaller parameter capacity, struggles with complex data or tasks requiring high precision, resulting in inferior performance across all metrics.

Furthermore, SeaThru-NeRF exhibits excellent performance with both synthetic and real datasets. This indicates that the larger model generalizes better to different data distributions, effectively modeling and predicting both low-noise synthetic data and more uncertain real data. This enhanced generalization ability also enables SeaThru-NeRF to maintain low error metrics under various experimental conditions.

It is worth noting that SeaThru-NeRF-lite was designed as a lightweight model to be used in resource-constrained environments. Therefore, despite its inferior performance compared to SeaThru-NeRF, it still offers unique advantages and application value in scenarios where computational resources and model performance must be balanced.

In summary, the superior performance of the SeaThru-NeRF model under various experimental conditions can be attributed to its larger model capacity, stronger feature representation ability, and better generalization performance. In practical applications, selecting the appropriate model architecture based on specific needs and resource availability is crucial for achieving optimal performance.

\subsubsection{Influence of datasets}

In the synthetic datasets, we found that the three AUSE metrics are generally higher than those in the real datasets. This can be analyzed in detail from the perspectives of data complexity, noise and data quality.

On the one hand, synthetic datasets are usually generated based on specific rules and models. Although they exhibit consistency and predictability, the simplified assumptions and lack of real-world complexity during the generation process may cause the model to overfit when dealing with synthetic datasets. This overfitting phenomenon can lead to increased prediction errors for the model on synthetic datasets, thereby raising the values of uncertainty estimation. In other words, the synthetic datasets may be overly idealized, failing to fully reflect the complexity and diversity of the real world, resulting in insufficient generalization ability of the model on these data, thus performing poorly in uncertainty estimation.

On the other hand, real datasets typically contain more noise and unpredictable factors, such as lighting changes, occlusions, and measurement errors. These factors can lead to larger errors for the model on real datasets but also force the model to learn more generalization capabilities during the training process, thereby improving the accuracy of uncertainty estimation. Although the data complexity and diversity in real datasets are higher, the model can exhibit lower AUSE values on these complex data through better generalization ability. The various uncertainties and randomness in real datasets, on the contrary, make the model more adaptable and robust when facing unknown data.

In short, the higher AUSE values in synthetic datasets compared to real datasets indicate that the model's performance on synthetic datasets is inferior to that on real datasets. This is mainly due to the simplified assumptions and lack of real-world complexity in synthetic datasets, leading to overfitting of the model on synthetic datasets, thus performing poorly in prediction error and uncertainty estimation. In contrast, the diversity and complexity of real datasets compel the model to possess better generalization capabilities, allowing it to more accurately estimate uncertainties and reduce prediction errors in complex data environments.

\subsubsection{Influence of errors}

Tables \ref{Tab.results}-\ref{iterations} document the results of our main experiments as well as the results of the ablation experiments. In these experiments, we calculated the AUSE values related to three types of errors (MSE, MAE, RMSE). The experimental results indicate that the AUSE\_MAE values are generally higher than the corresponding AUSE\_MSE and AUSE\_RMSE values under the same conditions.

Firstly, it is essential to understand the calculation methods and differences among AUSE\_MSE, AUSE\_MAE, and AUSE\_RMSE. AUSE\_MSE and AUSE\_RMSE measure the overall performance of model uncertainty using the MSE and RMSE, respectively. Since MSE and RMSE are more sensitive to larger errors, especially outliers, even a few large errors can significantly increase the overall value in these metrics. In contrast, AUSE\_MAE is based on MAE, treating each error value equally without excessive sensitivity to outliers.

The observation that AUSE\_MAE values are generally higher than AUSE\_MSE and AUSE\_RMSE may be due to the presence of some large errors in the actual error distribution. These large errors are amplified in the calculation of AUSE\_MSE and AUSE\_RMSE, resulting in lower overall values. However, AUSE\_MAE, being less sensitive to these large errors, shows relatively higher values. Thus, AUSE\_MAE provides a smoother measurement method when evaluating uncertainty quantification performance, avoiding the overemphasis on individual large errors.

Secondly, different datasets and models may generate different types of error distributions during training. For instance, in synthetic datasets, the model might more easily capture the overall features of the data, resulting in a more concentrated error distribution and consequently higher AUSE\_MAE values. For real datasets, the error distribution may be more dispersed with more outliers, causing AUSE\_MSE and AUSE\_RMSE to be elevated by these outliers, making AUSE\_MAE relatively higher.

Finally, the impact of experimental parameter settings must be considered. Variations in $M$, $\lambda$, and iterations will affect the degree of model training and data fitting. These changes directly influence the distribution and magnitude of errors, affecting the values of each metric. When evaluating the influence of these parameters, it is crucial to comprehensively consider the characteristics of different metrics and data distributions to fully understand the model's performance.

In conclusion, the phenomenon that AUSE\_MAE values are generally higher than AUSE\_MSE and AUSE\_RMSE suggests that the model's fitting to the data varies under different settings, reflecting the differences in sensitivity of these metrics to errors. In practical applications, selecting appropriate evaluation metrics should be based on specific application scenarios and data characteristics to more accurately measure model performance and uncertainty.

\begin{figure}
	\centering
        \makebox[\textwidth][c]{
	\includegraphics[width=1.3\columnwidth]{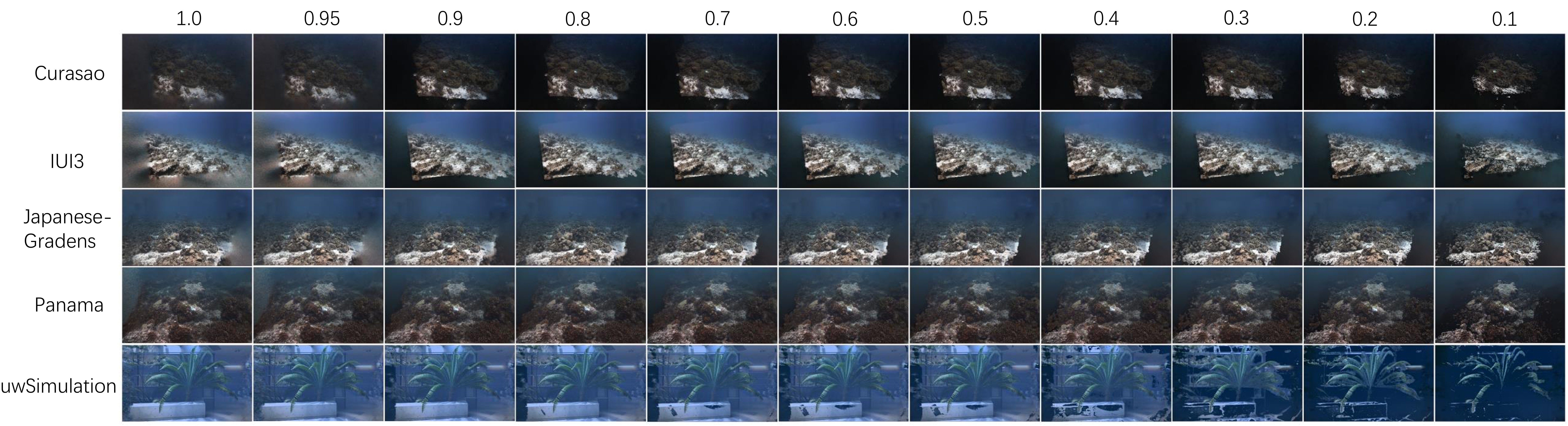}
        }
	\caption{By gradually lowering the uncertainty threshold, scenes with significant floating objects can be incrementally cleaned, ultimately yielding a clean target object image free of artifacts.}
	\label{FIG:6}
\end{figure}

\subsection{Applications: clean up}
In this work, we utilize Bayesian Laplace approximation to compute the covariance matrix $\Sigma$ for the perturbed field parameters $\boldsymbol{\omega}$. The diagonal elements of $\Sigma$ represent the uncertainty in each dimension. We render these diagonal elements as a new volumetric data $\mathcal{U}(\mathbf{x})$, where regions with larger $\mathcal{U}(\mathbf{x})$ values indicate higher uncertainty. These high uncertainty areas often correspond to artifacts in the rendering results. By applying a threshold to $\mathcal{U}(\mathbf{x})$, we can identify and remove these artifacts, resulting in clearer reconstruction outcomes (Fig. \ref{FIG:6}). Consequently, this method not only effectively quantifies the uncertainty in each dimension but also post-processes the reconstruction results.

\section{Conclusions}
In this paper, we introduce a spatial perturbation field $D_{\boldsymbol{\omega}}$ based on Bayes’ rays to quantify the spatial uncertainty of an underwater 3D reconstruction represented by neural radiance fields with scattering medium. This spatial perturbation field $\mathcal{D}$ makes a small perturbation to the input coordinates and inputs the perturbed coordinates into the SeaThru-NeRF to recalculate the colours and densities of object component to obtain the reconstruction results after the perturbation. The uncertainty of each spatial location is modelled by using the Laplace approximation method based on the difference between the original reconstruction results and the perturbed results. Moreover, by rendering the estimated spatial uncertainty field as an additional one colour channel, it is possible to visualise which regions of the whole scene have higher uncertainty. Furthermore, using the uncertainty field, we can remove artifacts from the rendering results of the underwater scene by simple thresholding. Numerical experiments show that our method can explicitly infer the spatial uncertainty of the model on both synthetic and real scenes, and exploit this uncertainty to improve the reconstruction quality. The method will benefit downstream tasks in ocean exploration and navigation, such as underwater reconnaissance and security surveillance, underwater navigation and localisation, and underwater infrastructure inspection. The current work has limitations: we assume that the light source is only from natural ambient light. However, when in deep-sea regions, the effects from artificial lighting and multiple scattering need to be considered due to poor visibility. We will investigate more diverse scenarios and medium parameter estimation methods in our future work.

\section*{CRediT authorship contribution statement}
H.L.: Conceptualization, Methodology, Software, Writing – original draft, Writing
– review \& editing, Funding acquisition. X.L.: Methodology, Investigation, Software, Writing – original draft, Writing – review \& editing, Data curation. Y.Q.: Conceptualization, Data curation, Supervision, Validation, Writing – original draft. J.D.: Investigation, Formal analysis, Resources. Z.M.: Formal analysis, Writing – review \& editing, Validation, Visualization. J.L.: Investigation, Validation, Visualization. L.C.: Conceptualization, Investigation, Project administration, Supervision.

\section*{Declaration of competing interest}
All authors certify that they have no affiliations with or involvement in any organization or entity with any financial interest or non-financial interest in the subject matter or materials discussed in this manuscript.

\section*{Data availability}
The data that support the fndings of this study are available from the authors upon reasonable request.

\section*{Acknowledgments}
This study was funded by the National Natural Science Foundation of China (No. 52274222).

\bibliography{sn-bibliography}

\end{document}